\documentclass[a4paper,12pt]{article}
\usepackage{amssymb,amsmath,cite,dsfont,bm}
\usepackage[scr]{rsfso}
\usepackage{graphicx}
\usepackage{color}
\usepackage{cases}
\usepackage{multicol}
\usepackage{wrapfig}
\IfFileExists{pictures._no}{\let\NOPISCURES\empty}{\let\NOPISCURES0%
\usepackage{xcolor}
\definecolor{red}{rgb}{1,0,0}
\usepackage{tikz}
\usetikzlibrary{decorations.text,calc,arrows.meta}
\usepackage[vcentermath]{youngtab}
\usepackage{tkz-euclide}
\usetkzobj{all}
\usepackage{pgfplots}
\usetikzlibrary{shapes}
\usetikzlibrary{decorations.shapes}
\usetikzlibrary{positioning}
}%

\def\Def{\stackrel{\:\mathrm{def}}{:=}}
\newcommand\ERR[1]{\textcolor{red}{#1}}
\newcommand{\br}[1]{\left(\!#1\!\right)}
\newcommand{\Br}[1]{\left[#1\right]}
\newcommand{\BR}[1]{\left\{#1\right\}}
\newcommand\INT{\int_{-\infty}^{+\infty}\!}
\newcommand\FSP[1]{\lfloor#1\rceil}

\def\sg{\kappa}
\def\state{\rho}
\def\starA{\mathscr{A}}
\def\starU{\mathscr{U}}
\def\pHi{\bm{\phi}}
\def\x{\bm x}
\def\Kv{\bm\xi}
\newcommand\KMS[2][\beta]{\state_{#1}(#2)}
\title{Key points for understanding of Unruh and Hawking effects}
\author{Peter Slavov\thanks{\tt petersl@phys.uni-sofia.bg}\\
{\it Sofia University ``St. Kliment Ohridski'',}\\
{\it Theoretical Physics Department,}}
\begin{document}
\maketitle
\begin{abstract}
The present paper stress on the distinction between the two idealizations~-- eternal (primordial) black hole and a black hole formed in a process of gravitation collapse.
Such a distinction is an essential condition for a better understanding of the relation between Unruh effect and Hawking effect.
The fundamental character of the discussions inside requires a preferred orientation to results of the local QFT in curved space-time.

\end{abstract}
\section{Introduction}
The meeting between the quantum physics and a space-time horizon is one of the most untrivial things in the physics. A serious challenge to common sense, a test about how we understand the basic physical concepts and notions like space-time, causality, locality, dynamics, evolution, quantization, temperature, information, entropy, energy, mass, particle.

In early years of General Relativity (GR) black hole (BH) has been an exotic feature of causality structure of the space-time. After obtaining Kerr's solution of Einstein--Hilbert's equations in 1963, the development of BH physics starts growing up as a separate discipline in gravitational science fruitful for many new results.
The Golden Age of this area of research was a highest pilotage in theoretical and mathematical physics. It consist of a few number of important theorems coming after a deep study of gravity solutions with symmetries and also of a brilliant (skillfully) defined system of basic notions leading to outstanding physical results (4 lows of BH mechanics \cite{BCH} and Smarr's formula \cite{Sm}) despite (regardless of) the difficulties in the physical interpretation and the conflict with common sense. To the last here it is an example. According to GR the gravitation field (the metric) does not have got its own 3D energy-momentum density (or 4D density of action), but it may have got such an integral characteristic like mass or as it is said in \cite{MTW} there is a gravitational energy but it cannot be localized. In 1987 Chandrasekhar wrote \cite[p.\,206]{P}:
\textsl{``This is the only instance we have of an exact description of a macroscopic object. Macroscopic objects, as we see them around us, are governed by a variety of forces, derived from a variety of approximations to a variety of physical theories. In contrast, the only elements in the construction of black holes are our basic concept of space and time. They are, thus, almost by definition, a perfect macroscopic object in the universe.''}

Nontopological characteristics (mass, angular momentum) of a BH by construction are not based on what BH consist of, but how it looks from outside. The mass of a BH as a notion does not derive from a distribution of matter inside it, but because Kommar's integral gives a result. Thus the mathematical theory of BH as if tell us that the questions about 'what BH consist of, what it composed of, how the matter inside is distributed and what kind interactions and processes happen?' are meaningless because they are irrelevant to the problem of the evolution of a BH, so we do not need such thing as a whole physical picture.

The rigorous mathematical result, the root of the whole this strange situation is so-called \textit{no hair theorem}. According to it, the only physical characteristics of a black hole in 4D space-time are mass $M$, angular momentum $J$ and electric charge $Q$. In the light of the no hair theorem the question about what happens behind the horizon after the collapse occurs seems somehow similar to the Feynman's father question about how the absorbed photon continues to live inside the atom. But there are two important differences. The collapse is treated classically. Besides, the most of the characteristics of the collapsing matter do not become the characteristic of the black hole.
According to Bekenstein conjecture, the BH must have got an entropy, proportional to the area of the horizon $S\sim A$. So if the generalized Smarr's formula will be rewritten as the main equation of the equilibrium thermodynamics,
\begin{equation}\label{Smarr}
dM=\frac\sg{8\pi} dA+\Omega dJ+\Phi dQ\;\stackrel{?}\longrightarrow \;dM=T_{\!bh}dS+(\dots)\,,
\end{equation}
then the temperature will be proportional to the surface gravity. If we suppose a quantum origin of the BH temperature, the unknowing constants have to contain Planck constant. To substantiate the equation of thermodynamics from first principles we need to derive the temperature or entropy or both from a fundamental theory.

Bekenstein conjecture actually set a requirement realizing the entropy as a \textit{model independent notion}. The modified Smarr's formula extends the requirement also to the temperature.

There are models deriving the entropy based on the classical physics. Also, there are models, deriving the temperature and BH radiation from a quantum tunneling through the horizon.
Hawking's decision \cite{H} is applying quantum field theory (QFT) on a curve space-time with a horizon. 

Sometimes Hawking's result\cite{H}  is taken as 'no rigorous' \cite{S}, or `heuristic' \cite{K3}. Nevertheless, its significance is huge so to add one more new metric to the zoo to obtain the expected Planck spectrum always is a praiseworthy work. There are a lot of publications irradiated by the BH quanta. Probably some authors do not render an account that their results are particular cases of already proved theorems.

Before any new confirmation of the expected precious spectral law, it is much more reasonable to check whether the serial new metric allows a self-consistent quantization at all.

\subsection{Global hyperbolicity as a restriction of space-time geometry}
First of all, the present status of a locally constructed quantum field theory (LQFT) requires a restriction of Lorentzian manifolds to a \textit{globally hyperbolic} such.
A Lorentzian manifold is called \textit{causal}, if it does not contain closed causal (time-like or light-like) curves. Globally hyperbolic manifolds are on the top of the hierarchy in the class of causal manifolds \cite{MS}. From the other side General Relativity permits even
\emph{totaly vicious} space-time as solutions of gravitational equation. G\"odel cosmos is the most popular example.

There are different definitions of global hyperbolicity. Robert Geroch has proved their equivalence  \cite{G} and also has determined the topology of a globally hyperbolic manifold $\mathbb R\times M$.
He also has proved the stability of the global hyperbolicity, which means that a small variation of the metric 
will preserve it.
For the needs of QFT, the definition related to a Cauchy surface or Cauchy region is the most suitable. Briefly speaking, globally hyperbolic is such a manifold, which whole is a Cauchy region for a hyperbolic differential equation.

A Lorentzian manifold can contain a Cauchy region even if it is not  global hyperbolic. The boundary of the global Cauchy region is known as \textit{global Cauchy horizon}. The word `global' means `not extendable to larger'.
Also, a globally hyperbolic space-time may contain subregion which is a Cauchy region itself or region of hyperbolicity. If a space-time contains a noncompact region of stationarity the last is a region of hyperbolicity. The simplest example is Minkowski space where a Rindler wedge is a subregion of hyperbolicity~-- a foliation of equal time open half-plains (nonglobal Cauchy surfaces).
\ifx\NOPISCURES\empty\framebox(200,200)[c]{\ERR{Picture}}\else
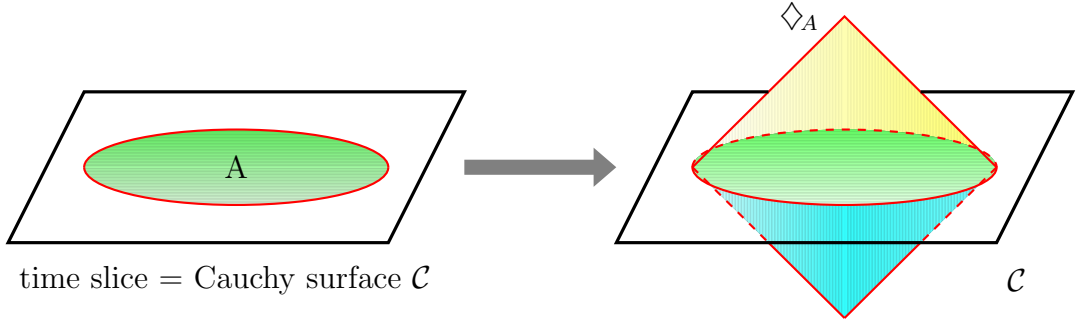
\begin{figure}[h!]
\begin{center}
\begin{tikzpicture}
	\draw[color=black, very thick] (-4, -1) -- (1, -1) -- (2, 1) -- (-3, 1) -- cycle;
	\fill[
	      top color=green!60,
	      bottom color=green!10,
	      shading=axis,
	      opacity=0.85
	      ]
	    (-1,0) circle (2cm and 0.5cm);
	    \draw[red, thick] (-3, 0) arc (180: 360: 2cm and 0.5cm) -- (1, 0);
	    \draw[red, thick] (-3, 0) arc (180: 0: 2cm and 0.5cm);
	
	\draw[line width=2mm,color=gray,>={Triangle[length=3mm,width=5mm]},->] (2, 0) -- (4, 0);
	
	\fill[
	      left color=yellow!0,
	      right color=yellow!60,
	      middle color=yellow!30,
	      shading=axis,
	      opacity=0.95
	    ]
	    (5, 0) -- (7, 2) -- (9, 0) arc (0: 0: 2cm and 0.5cm);
	\fill[
	      left color=cyan!10,
	      right color=cyan!50,
	      middle color=cyan!80,
	      shading=axis,
	      opacity=0.85
	    ]
	    (5, 0) -- (7, -2) -- (9, 0) arc (0: 0: 2cm and 0.5cm);
	\fill[
	      top color=green!60,
	      bottom color=green!10,
	      shading=axis,
	      opacity=1
	      ]
	    (7,0) circle (2cm and 0.5cm);
	\draw[red, thick] (5, 0) arc (180: 360: 2cm and 0.5cm) -- (7, 2) -- cycle;
	\draw[color=red, dashed, thick] (5, 0) arc (180: 0: 2cm and 0.5cm);
	\draw[color=red, dashed, thick] (5, 0) -- (6, -1);
	\draw[color=red, dashed, thick] (9, 0) -- (8, -1);
	\draw[color=red, thick] (6, -1) -- (7, -2);
	\draw[color=red, thick] (8, -1) -- (7, -2);
	
   \draw[color=black, very thick] (6, 1) --  (5, 1) -- (4, -1) -- (9, -1) -- (10, 1) -- (8, 1);
	\node at (-1, 0) {A};
	\node[right] at (-4, -1.5) {time slice $=$ Cauchy surface $\mathcal C$};
	\node[right] at (6, 2) {${\diamondsuit}_{\!A}$};
	\node[right] at (9, -1.5) {$ \mathcal C$};
    \end{tikzpicture}

    \end{center}
    \caption{Causal diamond associated with $A$.}
\label{fig:diamond}
\end{figure}%
\fi%

If cannot be found a Cauchy surface intersecting the whole manifold then we have a \textit{global Cauchy horizon}.

  The basic quantization (the determination of the canonical commutation relations (CCR) in space-time) for a linear scalar field  is set by the equation of motion (Klein--Gordon--Fock) plus equal time commutation relations ($\hbar=c=1$)
 \begin{equation}\label{CCR1}
 \vec x,\vec{\,x}' \in\Sigma_t\quad\left.\begin{array}{|l}
   {[\hat\phi(\vec x,t),\hat\phi(\!\vec{\,x}',t)]}=0\\
   {[\hat\pi(\vec x,t),\hat\pi(\!\vec{\,x}',t)]}=0\\
   {[\hat\phi(\vec x,t),\hat\pi(\!\vec{\,x}',t)]}=-i\delta^3(\vec x,\!\vec{\,x}')\hat1
  \end{array}\right\}\;\Rightarrow\;
[\hat\phi(\x),\hat\phi(\x')]=i{\cal D}(\x,\x')\hat1
\end{equation}
  on an arbitrary space-like hypersurface $\Sigma_t$.
The last becomes a Cauchy data for  Cauchy's boundary problem for a bisolution known as Pauli--Jordan's function: \ ${\cal D}(\x,\x')=G^{adv}(\x,\x')-G^{ret}(\x,\x')$.
This c-valued function propagates the quantization in the Cauchy region of space-time.

\begin{description}
 \item[\underline{Th:}]\textbf{ In a globally hyperbolic space-time there is an unique global retarded or advanced Green function\ $G^{adv/ret}(\x,\x')$} \cite{BF}.
 \end{description}
In the algebraic framework ${\cal D}(\x,\x')$ determines a \textit{symplectic structure} in the space of the test functions and the quantization means a map from a classical symplectic structure to a \textit{$*$-algebra} \cite[\!p.36]{BF}.
So \textbf{QFT is only possible in the region where Pauli--Jordan's function is uniquely determined.}

In BH a noncompact simply connected stationary region is a Cauchy region, bounded by a Killing horizon, so a Killing horizon always is a Cauchy horizon but not a global such when.
So if we try to construct QFT associated with such a region following the prescription of mode decomposition quantization (as it is made in Lorentzian coordinates), we have to know that there isn't a possibility to extend the quantization behind the horizon unless using some additional conditions.

On the level of hyperbolic differential equations, the existence of a nontrivial solution beyond the Cauchy horizon is not broken, but the uniqueness is. So additional principles we need for selection of a prefer continuation of ${\cal D}(\x,\x')$.
In that sense, the global hyperbolicity restriction rather has got a character of a reinsurance than a NoGo-theorem.

A possible construction for the propagation of the symplectic structure beyond the domain of dependence of the initial data surface is based on a \textit{self-adjoint extension of a differential operator} \cite{Sa}.


 Besides the standard topological notion of neighbourhood related to the smooth structure of a manifold, there is another one, related to the Lorentzian metric~-- the \emph{hyperbolic neighbourhood of a point}.
 If a time-like surface passes through the fixed point, the positive definite metric determines a topological, Hausdorff type neighbourhood on it, which may be considered as a Cauchy area for a hyperbolic differential equation.
 The isotropic geodesic passing through each boundary point form a local Cauchy horizon closes a restricted in time and space open region~-- the Cauchy region or the hyperbolic neighbourhood of a point (see the picture).
A local quantum algebra is uniquely determined in it by equal time commutation relation on the local Cauchy surface.

 Thus the hyperbolic neighbourhood of a point, called in the literature  \cite[\!\dots]{H0,F0,K1,K3,M,V} \textit{`double cone'} or \textit{`diamond region' } or also \textit{`open or causal diamond'}, is the natural geometrical notion of a causally completed neighbourhood $\diamondsuit$, associated to a local quantum algebra $\starA(\diamondsuit)$, built by a Cauchy boundary problem. The Haag's concept of locality realized as a net of local algebras restricted to a relativistic framework gains the statement about the proper way of covering of a Lorentzian manifold by a net of opened diamonds.

 Bernard Kay continues Haag's idea of locality introducing a \textit{principle of  \footnote{\ In algebraic QFT the restriction of a field to subregion is accomplished by the restriction of the support of a \textit{test function}. Kay denotes the test function by `$F$.'}\!F-locality} \cite{K1,K2} It declares that
 \textsl{every neighbourhood of any point in space-time has to contain an open diamond, so the restriction of the global field algebra to the diamond coincides with a separately built in it local algebra.}

 The globally hyperbolic space-times are fully compatible with the principle of $F$-locality.

There are many space-time models with BH that are not globally hyperbolic. For instance, Reissner--Nordstr\"om type solution in GR for some combinations of parameters permit a second \textit{Killing horizon}\footnote{\ It will be defined later.} that turns out be a Cauchy horizon. Since the internal Killing horizon covers a compact static region, according to a theorem \cite{MS}, such a region contains closed chronological curves, therefore it is unacceptable from a point of view of our basic concepts of evolution.

Hawking shows \cite{H1} that in space-times with a compactly generated Cauchy horizon QFT leads to a violation of \emph{the weak energy condition}.
(Hawking's chronological conjecture). Further development based on the algebraic QFT supports this conjecture by two NoGo theorems \cite{KRW}
According to them the space-times with a compactly generated Cauchy horizon violate F-locality, and the quantum stress--energy tensor is singular on such a horizon.
In simple words, there is not a self-consistent QFT on the class of metrics containing such type horizon, so it isn't possible to be constructed a time machine \cite{K3}.
\vspace{.2ex}





\subsection{Algebraic QFT}
The exceptional prediction power of  QFT as a constructive science (Lemb shift, magnetic moments, Casimir force as well as all great successes of particle collision physics up to nowadays) push back the important formal problems of QFT.
The arguments often used are based on conditions in witch the effective asymptotic of the theory are the dimensions of the laboratory, so the influence of space-time curvature is negligible even for the high accuracies quantum physics measurements realize.

Beyond whole  success of the concept of gauge field theory, group representations theory together with Wigner's concept of particle and vacuum, also quark model,  continual integrals, renormalization methods etc.,
fundamental questions like these remain:

How a locally formulated QFT can be compatible with highly nonlocal effects like quantum entanglement?
How to understand a concept of particles in the case when the symmetries are missing? All this things require a deeper understanding and a revision of the basic structure of QFT.



 The great legacy of the General Relativity, beyond the Einstein's equations, leads to the conviction that \textbf{every fundamental physical theory must be formulated in a covariant manner}.
The main obstacle for such a formulation is a theory with fixed Hilbert space and linear operators acting on it. For systems with finite degree of freedom, the representations of commutation relations are equivalent (Stone--von Neumann theorem) \cite{HW}, while the systems with an infinite degree of freedom suppose an infinite number inequivalent algebraic representations (infinite many inequivalent vacua). In Minkowski space, the Poincar\'e invariant state determines a preferred representation of quantum algebra with a spectral condition allowing a concept of particle and vacuum related to a mode decomposition quantization. In a curve space-time in general case, there isn't symmetry so there isn't such a preferred state, there isn't a preferred notion of positive frequency part of the solution of field equation. That's why there isn't a Hilbert space with an associated to it Fock space.
$${\cal F}({\cal H}):=\mathbb C\oplus{\cal H}\oplus({\cal H}\otimes_S{\cal H})\oplus\dots$$

An old program rule says: Before describing something new in a curve space-time first try to describe it in a noninertial reference frame in Minkowski space.
Fulling realises an alternative mode decomposition quantization \cite{F} attached to a curvilinear time symmetry where the quantum field is located in the double wedge region of Minkowski space. Unruh have demonstrated that Fulling--Rindler particles could be detected by a uniformly accelerated observer. So which is the right notion of a free particle 
from a viewpoint of uniformly accelerated observer?
In that way, the constructive physics offers new simple models as an alternative to Wigner's concept of particle as irreducible unitary representations of Poincar\'e group with a positive energy. Further development in this direction is Hawking effect. The cross point of all new schemes of quantization base on the fact that the frequency separated quantization uses only time symmetry. The problems formally formulated and solved by Parker, Fulling, Unruh and Hawking open a new window to a large-scale manifestation of QFT.

Although there isn't a perspective for experimental realizations in a near future, all these relatively simple constructive decisions catalyze a research of the formal structure of QFT.




\subsubsection{$*$-algebra, states and Hilbert spaces}
The locally formulated QFT on the language of $*$-algebras is that formulation which preserves the quantum mechanical paradigm of noncommutative observables eliminating at the same time the central position of Hilbert space, transferring it to the reserve.
From the course of quantum mechanics, we know that the linear operators in Hilbert space may form an associative algebra, where the composition of action plays a role of multiplication. Besides there is a relation $(\hat A\hat B)^\dag={\hat B}^\dag{\hat A}^\dag$.

Subtracting all these properties from the context of its operator realization in a fixed Hilbert space, it can be introduced an abstract associative algebra $\starA$ with unit $\mathds1$ and conjugation $*$, satisfying the following properties by definition for all $A, B \in\starA$:
\begin{enumerate}\itemsep=0ex
  \item $A^{**}:=(A^*)^*:=A$\quad\textit{ involution},
  \item $(aA+bB)^*:=\bar{a}A^*+\bar{b}B^*$\footnote{\ We shall use the standard in mathematics symbol for complex conjugation.} $\forall a,b\in\mathbb C$ -- \textit{anti-linearity } ,
  \item $(AB)^{*}:=B^*A^*$ \ \footnote{\ The symbol of multiplication traditionally is missing.}
\end{enumerate}

For the ability to calculate expectation value and statistical moments we need a notion of  \emph{state}, which by definition is a linear functional $\rho:\;\starA\to\mathbb C$ \ 
with the following properties:
\begin{enumerate}\itemsep=0ex
  \item\textit{real} -- $\rho(A^*):=\overline{\rho(A)}$\quad corollary: if $A=A^* \;\Rightarrow\;\rho(A)\in\mathbb R$
  \item\textit{positive defined} -- $\rho(A^*A)\geqslant0\;\;\forall\;A\in\starA.$
  \item\textit{normalized} -- $\rho(\mathds1):=1$
\end{enumerate}
As a corollary we have a natural definition of inner product as a symmetric positive defined sesquilinear form in $\starA$~-- $\langle A,B\rangle:=\rho(A^*B)$, but which is degenerate. So the algebra itself cannot be considered like a Hilbert space. But by a simple factorization procedure known as \textbf{G}elfand--\textbf{N}aimark--\textbf{S}egal construction, this obstacle can be overcome. For the pair $(\starA,\rho)$ there is
a natural homomorphism $\pi_\rho$ of $*$-algebra in a pre-Hilbert space\footnote{\ Complex linear space with a \textbf{non-degenerate} positive definite scalar product. For a completion to Hilbert space in sense of metricity, the additional conditions are required for the $*$-algebra and the state.} ${\cal H}_\rho$, which is a representation of $\starA$ \textit{(GNS-representation}) and a \textit{cyclic vector } $\Omega_{\!\rho}\!\in\!{\cal H}_\rho$ (algebraic or \textit{GNS-vacuum}). The index denote the dependency of the the three objects $(\pi_{\!\rho},{\cal H}_\rho,\Omega_{\!\rho})$, called \textit{GNS-triple}, on the choice of the algebraic state $\rho$.
This GNS-theorem, continue declaring the existence of an unique operator $U_{\!\rho'\!\rho}$ fulfilling the relation $\Omega_{\!\rho'}=U_{\!\rho'\!\rho}\Omega_{\!\rho}$ for arbitrary two states $\rho$ и $\rho'$ for a fixed algebra $\starA$.

In the algebraic framework the notions `state' and `vacuum' are often taken as synonyms due to the GNS-relation
$$\rho(A)=\langle\Omega_{\!\rho}|\,\pi_{\!\rho}(A)|\Omega_{\!\rho}\rangle\,.$$

\textit{Cyclicity}~-- the ability for generating the whole pre-Hilbert space $\pi_{\!\rho}(\starA)\Omega_\rho={\cal H}_\rho$~-- is that property which is the crossing point for Fock vacuum in QFT and GNS-vacuum. But the parallel finishes with this, since the GNS-vacuum, as a generic notion, is quite poor of properties. The state and GNS-vacuum are not attached to any kind of symmetry. They are analog of a coordinate system on a manifold. By reason of this, even when $\starA$ is specified as a particular quantum algebra, in general case the GNS-vacuum cannot be treated as a ground (lowest energy) state of the quantum field, so ${\cal H}_\rho$  cannot be associated to any space of excitations. That's why there is not a realization of a QF in terms of particles in the general case.
\textsl{``In addition, it is worth noting that the algebraic notion of states dispenses with the unphysical states in the Hilbert space that do not lie in the domain of the observables of the theory; vectors in a Hilbert space representation of the theory that do not lie in the domain of all  $\pi_\rho(A)$ do not define states in the algebraic sense.''} /R.\,Wald \cite{W3}/

The Riemannian manifold has got an important property of the existence of normal coordinates (not only in a point neighbourhood but in a neighbourhood of a geodesic curve). This makes it as a properly adapted mathematical structure for the formulation of General Relativity based on the principle of equivalence.
In the similar way, the GNS-triple is the core of  the success of the $*$-algebras language for the formulation of QFT in a covariant way, preserving the important role of Hilbert space. So GNS-triple, without superlative, could be called The Trinity of $*$-algebra.
\ \\

On the level of $*$- and $C^*$-algebra, before any reference to physics, we have certain kinds of states as a very primary notions. Some of them are directly defined as \textit{mixed state}~-- a \textit{convex} combination (mixture) of states~-- and \textit{pure} or \textit{extremal state}~-- an \textit{extremal convex combination} of states.


\underline{\textbf{Definitions:}}
%
\vspace{-4ex}
$$\begin{array}{|l}
    \state_1\neq\state_2\\
   \state=\lambda\state_1+(1\!-\!\lambda)\state_2\\
    \lambda\in[0,1]
  \end{array}\quad\Rightarrow\quad
  \left\{\begin{array}{cc}
   0<\lambda<1&\textit{mixed state}\\
   \lambda=0 \vee  \lambda=1&\textit{pure state}
  \end{array}\right.
$$

\underline{\textbf{Lema:}}\hfill pure state $\quad\Longleftrightarrow\quad$ irreducible (faithful) GNS representation.\hfill\phantom{Lema:}
\ \\

\underline{\textbf{Definition:}} \emph{Let $G$ is a group and $\alpha_g$ be an automorphism of $*$-algebra $\starA$ induced by $g\in G$.
A state $\state$ is called \textbf{$G$-invariant} if for all $A\in\starA$:}

\emph{a) $\state(\alpha_g(A))=\state(A)$\qquad or\qquad b) $\exists\;U_g^{-1}\!=U_g^\dag: \;\pi_\state(\alpha_g(A))=U_g\,\pi_\state(A)U_g^\dag$.}\vspace{2ex}

Other kinds of states are defined only by the properties of their GNS representations.

\underline{\textbf{Definitions:}} \textit{Two states are (unitary-\footnote{\ unitary equivalence = equivalence, \quad quasi-equivalence -- a notion for equivalence in a weaker sense \cite{HK}}, quasi-)equivalent if and only if their GNS representations are (unitary-, quasi-)equivalent.}

\underline{\textbf{Definition:}} \textit{Two GNS representations are unitary equivalent if for each $A\in\starA$ exists such an operator $U(A)$ \ so \ $\pi_{\!\rho'}(A)=U(A)\,\pi_{\!\rho}(A)\,U(A)^{-1}$.}

\underline{\textbf{Definition:}} \textit{Two GNS representations are quasi equivalent if they are realized as isomorphic von Neumann algebras.}



In an algebraic formulated QFT we get use to look at an observable as an object characterized by all statistical moments rather than by a spectrum of operator in a Hilbert space.
\ \\

\subsubsection{Local QFT and observable at a point}
To avoid some formal problems, an observable at a point is not a well-accepted notion in a LQFT \cite{W3,HW}.

Instead as an operator-valued function $\hat\phi(\x)$ or distribution on a fixed Hilbert space, the quantum field is defined as a \textit{smeared field} $\phi(f)$~-- a $C^*$-algebra-valued distribution, realized by the help of an ordinary c-valued \textit{test function} $f(\x)$, with a compact support  in a open bounded region ${\cal O}\subset M^4$.
$$\pHi(f)\Def\displaystyle\int_{\x\in M^4}\pHi(\x)f(\x)\,\mu(\x)\,,\quad\mu(\x):=\sqrt{-g(\x)}d^4x\vspace{1ex}\,,$$
The symbol $\phi(\x)$ does not have a separate meaning. \textbf{The \textit{smeared field} is the primary notion for local observable}.

So a local field is a member of a local algebra  $\phi(f)\in\starA(\cal O)$ and a set of quantum fields realises in \textit{a net of local algebras} $\bigcup_n\starA({\cal O}_n)$, where $\bigcup_n{\cal O}_n=M^4$. The definition of smeared field provide the axiomatic property of local algebras ${\cal O'}\subset{\cal O}\Leftrightarrow\starA({\cal O'})\subset\starA({\cal O})$. CCR (\ref{CCR1}) have to be rewritten in terms of smeared fields of $*$-algebra of the canonic commutation relations for free linear scalar field (CCR-algebra).
\begin{equation}\label{CCR2}
\begin{array}{l}
  \pHi(\bar f):=\pHi^*(f)\\
  \pHi((\Box-m^2)f)=0\;\;\forall f\in D(M^4)\\
 {[\pHi(f_1),\pHi(f_2)]}=i{\cal D}(f_1,f_2)\mathds1\,,
  \end{array}
\end{equation}
{\small$$\textrm{where }\quad
{\cal D}(f_1,f_2):=\int_{(\x,\x'\!)\in M^4\!\times M^4}\!f_1(\x){\cal D}(\x,\x')f_2(\x')\:\mu(\x)\!\wedge\!\mu(\x')
 $$}
 The above CCR (\ref{CCR2}) are compatible with the relativistic causality.

There are enough arguments against the notion `observable at a point' even in the framework of QFT with fixed Hilbert space \cite[\!p.3]{W3}. It's interesting to remark that Wightman is a preacher of the concept of smeared field \cite{Hh}. Nevertheless the symbol $\pHi(f)$ has remained as a mark of recognition of a research using the algebraic approach. Is this simply a tradition or something more? Deeper arguments against  `observable at a point' take roots in the original formulation of a \textit{quantum principle of locality} raised by Haag and Kastler \cite{HK}. More detail analysis can be found in \cite[\!sec.6]{Hh}.
 $$+\left.\begin{array}{|rl}
 1.&\textrm{no superselection rules}\\  
 2.&\textrm{ $*$-algebraic language}\\
 3.&\textrm{infinitely many inequivalent representations}
 \end{array}\right\}\;\;\longrightarrow \;\;\textrm{net of $C^*$-algebras}$$

 The root of the high complexity of quantum in compare to the classical physics is in an indispensability of the states as a part of the theory.
Haag, Narnhofer and Stein raise the \textit{Principle of Local Definiteness} (PLD) \cite{HNS} as a possibility to realize a state with physical meaning without knowledge of a reference state~-- ``\textsl{\dots for a small contractible neighborhood of a point the theory must fix the algebra uniquely in such a way that no superselection rules between the allowed ``partial states'' of such a small region exist.}''. After that, they translate this principle to the language of $*$-algebras~-- \textsl{``\dots the algebra of a compact, contractible region shall be isomorphic to an operator algebra with trivial center\footnote{\ $*$-algebra with a nontrivial center offers inequivalent GNS-representations. ``No superselection rules'' means equivalent GNS-representations.} on a Hilbert space and that only normal states\footnote{A state $\state$ is called normal, if for every monotone, increasing net of operators $H_\alpha$ with upper bound $H$, $\state(H_\alpha)$ converges to $\state(H)$.} of this algebra are allowed.''}
(see also \cite{H0} for a more detailed introduction to local QFT).
\begin{quote}\small
According to Haag's understanding, `no superselection rules for a local observable' is a more fundamental physical requirement than the chosen mathematical language. Since there is not a short path from a $*$-algebra to locally formulated QFT, in the contrast to many contemporary authors, Haag found `local QFT' as a more suitable name than `algebraic QFT'.
\end{quote}

The fact that every measurement pass in a small region with finite precision often declares as an argument for smeared observable.
But physically it is incorrect because outside the pure quantum measurements QFT also has to realize an interface between itself and GR or similar classical metric gravitational theory. We need a physical notion of state and also a concept of observable in point to identify Einstein tensor with an averaged stress-energy tensor.
\begin{quote}\small
In the QFT on the classical gravitation background (also known as quasi-classical approximation) the space-time geometry is treated as almost unaffected from vacuum fluctuations of the quantum field. But the influence of a quantum field on the space-time geometry no matter how weak it is, cannot be totaly neglected. It gives the effect of black hole evaporation.
\end{quote}
 A simple intersection of a sequence of smeared observables based on the primary algebraic structure led to a trivialization, because of PLD. To build an observable at a point we need a completion of algebras in a suitable topology which allows the appearance of unbounded operators \cite[\!p.3679]{BH}. The last is an additional structure, an upgrade. That's why we may say that \textbf{observable at a point appears as a secondary notion} in a higher place in the logical hierarchy of a locally constructed QFT.


\cite{BH} is a recommend useful guide 
with a lot of profound comments and references to subtle questions.

An important property of the algebra with a physical meaning is \textit{hyperfiniteness}~-- the ability to be well approximated by a finite dimensional $C^*$-algebra. The most important class of algebras possessing hyperfiniteness are\textit{ von Neumann's algebras} \cite{Hh}.
\ \

\subsubsection{States with physical meaning}
On the level of local quantum physics, we meet \ normal state, quasi-free state, ground state, Hadamard state, KMS-state,  etc.
A general restriction to normal states coming from the PLD represents the lowest level of requirement related to `physical meaning'.

An important class of states are \textit{quasi-free} or \textit{Gaussian states}, the states fully characterized by the two point correlation function.
 because they are relevant to the concept of particles.
The GNS construction for an arbitrary quasi-free state possesses a natural Fock space structure and vice versa \cite{KW}.

Two states are physical equivalent if and only if its physical content is the same.
The quasi-equivalent states are usually taken as corresponding to physical equivalent states.
Different criteria for quasi-equivalence of quasi-free states are found \cite{D}.
\begin{quote}\small
In the algebraic approach an exponential realization of CCR (by the set $\bm W(f):=\exp(i\pHi(f))$) is preferred instead (\ref{CCR2}) as a $C^*$-algebra with a bounded norm, also known as \textit{Weyl algebra}.
 One can obtain its explicit form using (\ref{CCR2}) and Baker--Campdell--Hausdorff formula. \vspace{2ex}
\end{quote}

Physical meaningless states are not those whose physical interpretation face difficulties, but those who make the formal structure of theory not self-consistent. For a further precision of the mathematical structure of QFT in curved space-time we have to raise an additional criterion about physical state.\vspace{0.3em}


Another important class is \textit{Hadamard states} $\state_H$. In the case of CCR-algebra, they are states whose symmetric part of two-point distribution is identical with Hadamard’s fundamental solution of a homogeneous wave equation \cite{DB},
\begin{equation}\label{H-state}
  {\cal W}^S(\x,\x'):=\state_H(\BR{\pHi(\x),\pHi(\x'})\Def
  \frac{u(\x,\x')}{s(\x,\x')}+v(\x,\x')\ln\!|s(\x,\x')|+w(\x,\x')\,.
\end{equation}
  Here $s(\x,\x')$ is the square of geodesic distance between the points $\x$ and $\x'$, and $u(\x,\x')$, $v(\x,\x')$ и $w(\x,\x')$ are smooth. The information about the state is contained in $w(\x,\x')$, while $u(\x,\x')$ and $v(\x,\x')$ are state-independent functions.
Hadamard bi-distribution has got an appropriate singularity structure \cite{FNW}, providing regularizability of stress-energy tensor in a covariant manner.
It turns out that Hadamard criterium is sufficient to accomplish a self-consistent completely covariant subtracting procedure for building the all Wick powers \cite{W3}.

How many distinguished Hadamard states exist? Verch proved that all quasi-free Hadamard states for Klein-Gordon field on a globally hyperbolic space-time are quasi-equivalent \cite{V}. He also proved, that they turn out to be normal states.

So \textbf{Hadamard state is the name of algebraic state with physical meaning}.
It is an additional axiom of a linear scalar field theory.

Just the Hadamard condition (\ref{H-state}) is the second criterion in \cite{KRW,K3} supporting the chronological conjecture.

\ \\

\section{Eternal black holes}
\subsection{Briefly about the space-time geometry}
In coordinates attached to the symmetry the horizon present as a coordinate singularity. The joint formulas for the metric, the \textit{Killing vector} and the \textit{surface gravity} for both cases~-- Rindler space-time and the space-time with a static, spherical-symmetric BH looks like this:
\begin{equation}
  ds^2=-g_{00}(x^1)(dx^0)^2+g_{11}(x^1)(dx^1)^2+g_{ij}dx^idx^j\,,\;\;i,j=2,3\:,\quad g_{ij,0}=0\,,\;\underline{g_{00}(x_{H})=0}\,;
\end{equation}
\begin{equation}
  \Kv=\partial_0\;,\qquad\sg:=\left.\frac{\sqrt{-g_{00}}_{,1}}{\sqrt{g_{11}}}\right|_{x^1=x_{H}}\,.
\end{equation}
To demonstrate the parallel between Unruh effect in Rindler wedge and particle creation in Schwarz\-schild--Kruskal like geometry we will prefer coordinates smoothly covering the horizons in both cases.
\begin{subequations}
\begin{numcases}{\{\mathbf g,\Kv: \pounds_{\!\Kv\,}\mathbf{g}=0\}\rightarrow\;\Kv=\sg(V\partial_V\!-\!U\partial_U)\,,\;\;}
ds^2=-dUdV+dx^2+dy^2\;,\qquad\sg=a&\label{M}\\
ds^2=-F(UV)\,dUdV+H(UV)\,d\Omega^2&\label{K}
\end{numcases}
\end{subequations}%
where $a>0$, $\sg>0$ are constants, and $F(UV)>0, H(UV)>0 \;\forall\,U,V$. Also can be added $F(0)=1$.
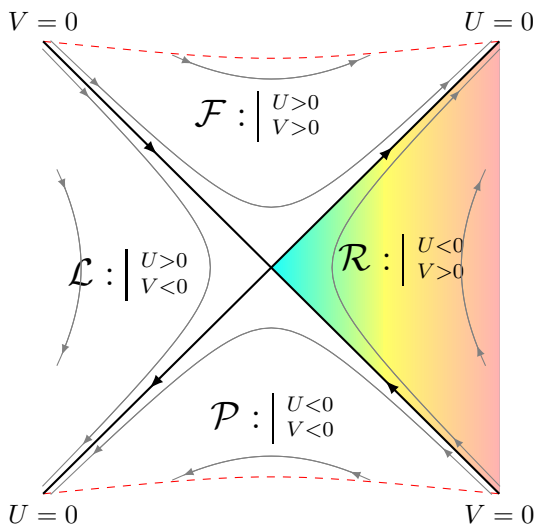
\begin{wrapfigure}{l}{0.4\textwidth}%
\ifx\NOPISCURES\empty\framebox(200,200)[c]{\ERR{Picture}}\else
\newcommand\Region[3]{\large${\cal #1}:{\scriptsize\begin{array}{|c}U{#2}0\\V{#3}0\end{array}}$}%
\def\hyperbola#1#2{
 \draw[>->,color=gray]plot[domain=-#1:#1]({#2*cosh(.9*\x)},{#2*sinh(.9*\x)});\draw[color=gray] plot[domain=-#1:#1]({#2*cosh(\x)},{#2*sinh(\x)});%
 \draw [>->,color=gray]plot[domain=-#1:#1]({-#2*cosh(.9*\x)},{-#2*sinh(.9*\x)});\draw[color=gray] plot[domain=-#1:#1]({-#2*cosh(\x)},{-#2*sinh(\x)});%
 \draw[>->,color=gray]plot[domain=-#1:#1]({#2*sinh(.9*\x)},{#2*cosh(.9*\x)}); \draw[color=gray] plot[domain=-#1:#1]({#2*sinh(\x)},{#2*cosh(\x)});%
 \draw[>->,color=gray]plot[domain=-#1:#1]({-#2*sinh(.9*\x)},{-#2*cosh(.9*\x)});\draw[color=gray] plot[domain=-#1:#1]({-#2*sinh(\x)},{-#2*cosh(\x)});%
 }
%
\begin{tikzpicture}[>=latex]
\fill[left color=cyan!90, right color=red!30, middle color=yellow!60, shading=axis, opacity=1] (3, -3) -- (0,0) -- (3, 3) ;
\draw[color=red,dashed] (-3,3) .. controls (0,2.7) .. (3,3);\draw[color=red,dashed] (-3,-3) .. controls (0,-2.7) .. (3,-3);
\draw[color=black, thick] (-3, -3) -- (3, 3);\draw[color=black, thick] (-3, 3) -- (3, -3);
\draw[->,thick] (1.5,1.5)--(1.6,1.6);\draw[->,thick] (-1.5,-1.5)--(-1.6,-1.6);\draw[->,thick] (-1.6,1.6)--(-1.5,1.5);\draw[->,thick] (1.6,-1.6)--(1.5,-1.5);
\node at (1.8,0.1) {\Region{R}<>};\node at (-1.8,-0.1) {\Region{L}><};\node at (-0.1,2) {\Region{F}>>};\node at (0.1,-2) {\Region{P}<<};
\node[above] at (3,3) {\footnotesize$U=0$};\node[below] at (3,-3) {\footnotesize$V=0$};\node[below] at (-3,-3) {\footnotesize$U=0$};\node[above] at (-3,3) {\footnotesize$V=0$};
\hyperbola{2}{0.8}\hyperbola{0.5}{2.5}
\end{tikzpicture}
\fi
\end{wrapfigure}%

In the case of Minkowski space-time (\ref{M}) $\Kv$ is the Lorentzian boost, $U$ and $V$~-- the Lorentzian isotropic coordinates, $a$\footnote{\ In SI $\sg [m^{-1}]$ and if $a\, [m.s^{-2}]$, then $a$ in  (\ref{M}) has to be replaced by $a/c^2\:[m^{-1}] $.}~--  the value of 4D acceleration. In the case of static spherically symmetric curved space-time (\ref{K}) $U$, $V$ are the \textit{Kruskal-like}\footnote{\ The cooordinates constructed in the same way as it is made for Schwarz\-schild--Kruskal geometry.} isotropic coordinates. 
In a curve space-time, it is also possible $\sg=0$. Then Kruskal-like coordinates are not possible.

In such coordinates, the horizon can be found as the places where the Killing vector field becomes a null vector.
\begin{subequations}\begin{numcases}{\mathbf g(\Kv,\Kv)=}
a^2UV\\
\sg^2UVF(UV)
\end{numcases}\end{subequations}
The academic name of 3D surface corresponding to the lines $UV=0$, often use by R. Wald is \textit{bifurcation Killing horizon}. For our further purposes, most benefit association of `bifurcation' will be `time reversible geometry'. The same about `maximally extended'. The coordinates $(U,V)\!=\!(0,0)$ \ $\Kv=0$ describe the unmovable points with respect to the isometry of our interest. In 4-dimensional case they correspond to 2D surfaces called \textit{bifurcation surface}. In the case (\ref{M}) it is the edge of the Rindler wedge, in the case (\ref{K})~-- the bifurcation 2-sphere.
The integral curve of $\Kv$, for which $\mathbf g(\Kv,\Kv)=-1$ the group parameter coincides with the canonical affine parameter, so $\Kv$ becomes 4-velocity on it. In the case of Minkowski space, the square of the distance from the horizon to such a curve is $UV=-a^{-2}$.

In the case (\ref{K}), U- and V-lines are light-like (geodesic) curves only on the horizon, but it is not important for our purposes. Also they may not cover the whole space-time. When they can, the manifold is globally hyperbolic.

If the metric tends to Minkowski type when $UV\to-\infty$, such a space-time is called \textit{asymptotically flat}.
\begin{center}
*\quad *\quad *
\end{center}

\textbf{Hawking effect has two sides~-- Hawking temperature and Hawking radiation. These two sides have to be distinguished, since they may correspond to different idealizations of space-time.}
Mixing the pictures corresponding to these idealizations may lead to a confusion (misunderstanding).

In Hawking's original paper \cite{H} BH is considered as formed by gravitational collapse. The main reason to put the stress on the distinction, related to the models of eternal BH and formed in a certain moment such,
originate in the existence of a significant number of publications with restricted considerations to the space-times with a perfect symmetry to derive the Planck spectral formula and to use it as an argument in favour of radiation.
For the inexpert reader, it may lead to misunderstanding, because a question about the role of gravitational collapse in the existing of the Hawking temperature remains open. A formal manipulation with Bogolyubov transformation as well other skilful methods of calculation as analytical continuations \cite{DR}, Wick rotation [?], asymptotic behaviour of Wightman function near the horizon \cite{HNS,Hs,M}, either may increase the confusion or may to lead us to an understanding that Hawking's temperature does not require a process of collapse, which it is not far from the truth.
Such a statement raises the question:
On what grounds the notion `Hawking temperature' come into use out of the context of the original paper \cite{H}? A bit later Hartle and Hawking offer an alternative approach describing the particle creation \cite{HH}, but the idealization is changed to a space-time with extended Schwarzschild geometry. One can see the conformal diagram there. After that `Hawking temperature' has become applicable to both idealization in various metrics with a horizon.
If the leading scientist changes the idealization preserving the interpretation of the result, what could expect from all the rest? Here it is the main part of the answer.

What is important for one or other method of obtaining how we get to account which idealization we deal with.



\subsection{KMS-state -- the adequate notion, characterizing by a temperature}
The mixture of the two idealization is also related to a mixture between particle creation and radiation.
\begin{itemize}\bf
\item The temperature implies a possibility for particle creation, but this does not mean radiation (emission) by all means. The question is: Where is the quant born?
\end{itemize}
One of the reasons of making a parallel between particle creation in Rindler wedge and in the BH exterior to achieve a clear understanding about the upper distinction.
I doubt if there is a uniform viewpoint among competent college about the importance of BH formation for the existing of radiation.

However strange it seems from a first look, the notion `temperature' can be suitable without a requirement of a process of BH evaporation, even without supposing any radiation.
The main contribution to this comes from the algebraic QFT, which puts the stress on the notion \textit{KMS-state}. Robert Wald uses the notion \textit{`Hartle--Hawking state'}, as KMS-state characterized by Hawking temperature not only for the case of extended Schwarzschild--Kruskal space but for space-times with a horizon and a perfect time symmetry~-- space-time with a bifurcation Killing horizon.\vspace{.8ex}

The evolution of the system with a Hamiltonian $\hat H=\hat H^*$ in Heisenberg representation can be given by the following algebraic automorphism
\begin{equation}\label{AlgEv}
\alpha_t(A):=e^{i{\hat H}t}Ae^{-i{\hat H}t}\quad\forall A\in\starU\,.
\end{equation}
In quantum statistical physics which operates with a finite-dimensional Hilbert space, the equilibrium thermodynamic state is a mixed state
 $$\state_\beta(A)=\mathrm{Tr}(\hat\rho_\beta A)$$
 with a density matrix given by Gibbs prescription
\begin{equation}\label{Gibbs}
   \hat\rho_\beta\Def\frac{e^{-\beta\hat H}}{Z(\beta)}\;,\quad  Z(\beta)\Def\mathrm{Tr}\,e^{-\beta\hat H}
\end{equation}
$Z(\beta)$ plays a role of statistical sum of Gibbs'  ensemble providing the property $\mathrm{Tr}\,\hat\rho_{\beta}\!=\!1$
and $\beta:=1/k_BT$ is an inverse temperature.

Now let's define a correlation function $F_\beta(A,B;t):=\KMS{B\alpha_t(A)}$ also supposing a complex analytical extension $F_\beta(A,B;z)$. Taking into account the cyclicity of the trace, it follows from (\ref{Gibbs}) $F_\beta(A,B;t\!+\!i\beta)=\KMS{\alpha_t(A)B}$.

This trace property has appeared for first time in Kubo's paper. Martin and Schwinger have used it only as a mathematical trick for obtaining equilibrium Green function.

If we forget (\ref{Gibbs}) and take the last equality by a definition instead as a consequence of (\ref{Gibbs}), then we go to the following

\underline{\textbf{Definition:}}\cite{KoS} \emph{The state $\state_\beta$ on an algebraic dynamical system $(\starU,\alpha_t)$ is called a $\beta$-KMS-state for $\beta>0$,  when it satisfies the KMS-condition at inverse temperature $\beta$, i.\,e. when for all operators $A,B\in\starU$ there is a holomorphic function on the strip $S_\beta:=\mathbb R\times i(0,\beta)\subset\mathbb C$ with a bounded, continuous extension to $\overline{S_\beta}$, such that
\begin{equation}\label{KMS1}
    F_\beta(A,B;t):=\KMS{B\alpha_t(A)},\quad F_\beta(A,B;t\!+\!i\beta):=\KMS{\alpha_t(A)B}\;\footnote{\ Sometimes the KMS-condition is typed in the form $\KMS{\alpha_t(A)B}=\KMS{B\alpha_{t+i\beta}(A)}$, but the compex continuation $\alpha_{t+i\beta}(A)=e^{(it-\beta)H}Ae^{(-it+\beta)H}$ is incorrect outside the correlation function, bacause $e^{\beta H}$ is an inbounded operator.}
\end{equation}}

In the pioneer work \cite{HHW} the notion `equilibrium state' is expanded and generalised for infinite volume systems. Haag, Hugenholtz and Winnik proved that the KMS-condition (\ref{KMS1}) can also be successfully applied in the case when the exponential operator in (\ref{AlgEv}) does not belong to the \textit{trace-class}, i.\,e. the definition of the Gibbs state (\ref{Gibbs}) does not make sense.

The research, made in an algebraic language,
has got a general character beyond the context of the relativistic or nonrelativistic physics. It based only on the following key assumptions:
 $C^*$-algebra has to allow a special GNS-representation known as \textit{von Neumann algebra} as well as the generator of the evolution $\hat H$ must be an operator with a positive spectrum.

This is an amazing example in theoretical physics when a particular property of one construction becomes its generalization. (In \cite{H0} it is shown the way back from KMS-state to Gibbs' ensemble  (\ref{Gibbs})).

The result is reported on the conference in Baton Rouge. On the same workshop, Minoru Tomita presents his ideas in the raw \cite{T} (historical reference \cite{Sr}), later reaching a completion by a collaboration of Masamichi Takesaki, systemized this kind of research as \emph{Theory of modular von Neumann algebras} or briefly called \emph{Tomita–Takesaki's modular theory} \cite{TT}.

\textsl{``\dots Tomita and Takesaki studied von Neumann algebras for which there existed a vector which is both cyclic and separating for the algebra. They found that such a vector (or rather the corresponding state) defines a distinguished one-parameter automorphism group for the algebra with some remarkable properties and a conjugation mapping of the algebra on its commutant. This group of modular automorphisms plays also an important role in physics. For instance, the extension of Gibbs’ cha\-rac\-teri\-zation of thermal equilibrium states to an infinitely extended medium is equivalent to the statement that \textbf{equilibrium is described by any state whose modular group is some one-parameter subgroup of time translations and (global) gauge transformations}\footnote{\ The emphasizing comes from me.}.''}\quad/Buholz, Haag \cite{BH}/

 At the times when Hawking's paper appeared, without any connection with it, nor with Unruh's work \cite{U}, Bisognano and Wichmann studied von Neumann algebra on Minkowski space as an application of the TT-theory and published what was later known as Bisognano--Wichmann's (BW) theorem \cite{BW1}. The content of the paper is cited approximately like this: \textsl{The algebraic state of the complete quantum algebra (or net of local algebras) becomes KMS-state after the restriction of full algebra to a subalgebra over the right Rindler wedge.} Or even in much more brief manner: \textsl{``Minkowski vacuum looks like KMS-state with Unruh temperature for uniformly accelerated observer.''}
But everyone can check, there is not any statement similar to this in whole the paper as well as in its generalization \cite{BW2}. Bisognano and Wichmann found that the modular group of a field von Neumann algebra is an algebraic realization of a one-parametric group of isometry preserving the Rindler wedge. The Killing vector (Lorentzian boost)  (\ref{M}) in it plays the role of a generator of evolution (quantized effective Hamiltonian).

A few years have been needed to pass BW theorem to attain importance in physics. To fall into a focus of attention for a then mathematical physicist, there is a need of a competent person that could translate such an abstract mathematical result to the appropriate physical notions. These merits usually are omitted in citations, but for s.o. who deal with a historical review 
it will worth be mentioned. In \cite[p.226]{HNS} can be found such a physical translation.
Sewell is usually cited as the first author who pointed out a connection between KMS-state, Hawking effect and BW theorem \cite{S}.
But the connection there based on a parallel between the temperature as a periodicity in an imaginary time, used for description of stochastic processes and the periodicity obtained in \cite{BW1}, rather than as an intrinsic structure encoded in TT-theory\footnote{\ In TT-theory the KMS-condition is encoded in the analytical continuation of a modular operator to an unitary such.}. The papers \cite{H,HH}, related to completely different idealizations of space-time, are cited en bloc in the context of KMS-state, although the only \cite{HH} is compatible with this algebraic notion. The message in \cite{S} is the importance of the algebraic framework for a rigorous treatment of Hawking effect.

The geometry of maximally extended Kruskal space-time and a large class of metrics are similar to maximally extended Rindler wedge~-- the external regions are analog to Rindler wedges, the Killing vectors (\ref{M}) и (\ref{K}) look similar, the quantum field in the exterior could probably be treated as a thermodynamic system with a infinite volume. So the natural question appears: Are there any obstacles to adapt Bisognano--Wichmann's result to the case of curved space-time with a bifurcation Killing horizon?

It turns out there are two independent conditions the global isometric invariant algebraic state has to satisfy:
\begin{enumerate}\itemsep=0ex
  \item It has to be a \textit{Hadamard state}. (A general requirement for self-consistency of QFT)
  \item The GNS-representation has to realize a von Neumann algebra with a modular structure. (A specific requirement for existing of a global KMS-state.)
\end{enumerate}
Both are fulfilled for Minkowski space. The two-point distribution for Minkowski state (vacuum)  becomes
\begin{equation}
{\cal W}(\x,\x')=\frac1{2\pi^2(x-x')^2}\,,
\end{equation}
where $x$ and $x'$ are Lorentzian coordinates.


The second condition is also satisfied for Minkowski state~-- \textit{Reeh--Schlieder's theorem} \cite{RS,H0}. (BW have referred to this theorem in their research.)

Bernard Kay proves that Schwarzschild geometry provides Reeh--Schlieder's conditions \cite{K0}. At the sight of this 
he considers a field algebra over the unity of left and right static regions (a BH analog of Fullung--Rindler quantization \cite{F}), he derives, that 
a primary pure state (pure vacuum) \textit{termalizes} to a KMS-state (KMS-vacuum) characterized by temperature
$$T_H=\frac\sg{2\pi}=\frac1{8\pi M}\,,$$
when the primary algebra is restricted to a subalgebra over the left or right static region.

It is proved later, that an arbitrary stationary space-time with a Killing horizon satisfies Reeh--Schlieder's conditions \cite{Str}.

Approximately in that time, TT-theory becomes an affirmative tool for research in mathematical physics. There is a nice review with references to open problems \cite{B}.

To emphasize the strong connection between a locally formulated free field quantum theory and the TT-theory, B. Schroer \cite{FS} introduces the term \textit{`modular localization'}, which gains an officiality due to Brunetti, Guido and Longo \cite{BGL}.

One of the most important results after Hawking's discovery is the theorems of the uniqueness \cite{KW}. Wald and Kay proved that an isometric invariant quasi-free Hadamard state does not exist for Ker and also for Schwarzschild--deSitter gravitational solution\cite{KW}. It also means that there isn't KMS state for these cases. Both class metrics characterized by an infinite number of bifurcation horizons. This achievement of the algebraic QFT demonstrates a \textbf{lack of a smooth transition from the idealization of entire BH to the idealization of formed BH}.\vspace{2ex}

For every two elements $A,B\in\starA({\cal R})$, where ${\cal R}$ is the region of localization,  KMS-state  (\ref{KMS1}) offers appropriate working formulae  in terms of spectral decomposition
\begin{equation}\label{KMS2}
 \KMS{AB}=\frac1{2\pi}\!\!\INT\!\frac1{1\pm e^{\beta\omega}}\!\INT\!\!\KMS{\underbrace{\alpha_t(A)B\pm B\alpha_t(A)}_{\textrm{c-valued function}\times\mathds1}}\,e^{-i\omega t}dt\,d\omega\,,
\end{equation}
The derivation of (\ref{KMS2}) uses a proper combination of pieces of the definition (\ref{KMS1}) in such a way to be obtained a state independent expression $\KMS\dots\!$  in the right sight.
 For linear field theories `$-$' corresponds to the case, when the algebra is set by commutator (it can be met in \cite{HNS}), whereas `$+$'%
~-- when algebra is set by anti-commutator. (I don't see anything against a similar formula for a more complicated case like $C^*$-superalgebras if only the dynamical system permits a $\beta$-KMS-state.)

In Unruh mode decomposition\footnote{\ Unruh modes = normal modes in BH exterior or Rindler wedge} ($\Kv f_{\pm\omega}=\pm i\omega f_{\pm\omega}$) near the horizon the quantized solution of the scalar field equation takes the form
$$\phi(U,V)=\int_0^{+\infty}\frac{d\omega}{\sqrt{2\pi}\sqrt{2\omega}}(f_{\omega}{\hat a}_{\omega}+\overline{f_{\omega}}{\hat a}_{\omega}^\dag+f_{\!-\omega}{\hat a}_{-\omega}+\overline{f_{\!-\omega}\!\!}\;{\hat a}_{-\omega}^\dag)\,,
\;\;\begin{array}{|l}
 U<0\\
 V>0
    \end{array},
\;\;\begin{array}{|c}
    f_{\omega}\approx(-\sg U)^{i\frac\omega\sg}\\
    f_{-\omega}\approx(\sg V)^{-i\frac\omega\sg}
    \end{array}
\,,$$
$$\textrm{where }\quad[{\hat a}_{\omega},{\hat a}_{-\omega'}]=0\,,\;\;[{\hat a}_{\omega},{\hat a}_{-\omega'}^\dag]=0\,,\;\;[{\hat a}_{\pm\omega},{\hat a}_{\pm\omega'}^\dag]=\delta(\omega\!-\!\omega')\,.$$
If $A$ and $B$ are creation and annihilation operators in the left side we have an expectation value for a total number of quanta with energy $\omega$  in $\beta$-KMS-state (with an inverse temperature $\beta$).

$$\Kv\;\;\stackrel{Q}\longrightarrow\;\;\hat H\;,\textrm{ where } [\hat H,{\hat a}_\omega^\dag]=\omega{\hat a}_\omega^\dag\;\;\Rightarrow\;\;\alpha_t({\hat a}_\omega^\dag)=e^{i\hat Ht}\,{\hat a}_\omega^\dag\,e^{-i\hat Ht}=e^{i\omega t}\,{\hat a}_\omega^\dag$$
\begin{equation}\label{KMS3}
N_\beta(\omega):=\KMS{{\hat a}_\omega^\dag{\hat a}_\omega}=\frac1{2\pi}\!\!\INT\!\frac1{1\!-\!e^{\beta\omega'}}\!\INT\!\!\KMS{\,\Br{{\hat a}_\omega^\dag,{\hat a}_\omega}\,}\,e^{i(\omega-\omega')t}dt\,d\omega'=
\frac{\delta(0)}{e^{\beta\omega}\!-\!1}
\end{equation}

The infinite number of quanta is related to the infinite time of the detector response. According to Haag's assumption, $N_\beta(\omega)$ cannot be treated as a local observable, but for a case of the detector with a finite time response, it can. 

If we set fields~-- $A:=\pHi(\x)$, $B:=\pHi(\x')$, we derive $\beta$-KMS-Wightman function $W_{\!\beta}(\x,\x')$.

In standard methods obtaining (\ref{KMS3}) (Bogolyubov transformations, analytical continuations in a neighborhood around the horizon,  etc.) the bosonic (or the fermionic) factor are a consequence of structures of the field equation. Very often this kind of derivations also refers to the asymptotical flatness of space-time. (I don't see any other important role of asymptotical flatness of space-time except a prefer notion of surface gravity.)

While in the contrast to this, the bosonic (or the fermionic)  factor appears at the very beginning in an outstanding model independent way~-- the formula (\ref{KMS2}) is a generic algebraic result not containing even relativistic physics. A particular field algebra is a further specialization.
KMS-state give the best answer to the question: why the different metrics lead to well known statistical spectrum? Only in this context, we have a strict physical notion of temperature by construction. The temperature appearing in the context of vacuum-vacuum transformation is only something that can be recognized by a Planckian spectrum, but how to say if it is related to an equilibrium process or state.

In the large review devoted to Unruh effect \cite{CHM}, the calculations are based on vacuum-vacuum transformation. When the authors discuss BW theorem they take a definition for KMS-state in terms of operators from \cite{K0}, which is closer to TT-theory and equivalent to (\ref{KMS1}). But this definition direct to the longer path of work in the frame of vectors in Hilbert space. That's why many containing Gamma-functions expressions are met there.

Why the path to the formula (\ref{KMS3}) seems so short? Because the much longer part of the path has already passed by theorems in the algebraic approach.

\textbf{The role of the algebraic framework in the cases of Unruh and Hawking effects is  a shifting  the focus of the attention from the inequivalent vacua to the field localization.}

Outside the $*$-algebraic framewirk there is a question about the meaning of vacuum-vacuum transformation.
If s.\,o. prefers this approach in spite of all, KMS would be a firm ground of understanding that \textbf{the first and the second vacuum are not in-going and out-going vacuum for the present level of idealization}.
 In the same way, if we describe a uniformly moving material point from two different reference frames the deference of the velocities we will obtain will not able to be interpreted as changes of the state of motion. But in the contrast to classical physics, where we don't need of second reference frame, in quantum BH physics we need a global vacuum and a vacuum attached to a causality connected stationary region to account the field localization in it.

Sometimes the vacuum-vacuum consideration may lead to the opposite statements about the real status of Unruh effect \cite{NFKMB}. Although at least one of the authors seems to be a $*$-algebraist, the final conclusion (p.025004-19) is that Unruh effect does not have a place in QFT and BW theorem is not relevant to Unruh effect. It seems to me this team faces difficulties of methodological character, but the boss of the team probably prejudiced against Unruh effect \cite{}. Their conclusion base on their view that a local observer has a duty to prepare a global pure state. But he cannot carry out his duty because of the restriction of the region of motion. And without the ability to prepare a global state, he cannot recognize a KMS-state. Or in a more algebraic language it sounds like this:
The observer has to know (by measurements) the complete algebra to have an ability to know a localized such. I don't agree with this kind of duty. Similar kind of duty cannot be realized even in classical BH physics. Can I consider space-time metric beyond the horizon without the possibility to perform a complete triangulation in whole manifold?

Methodologically it is important to distinguish using something as a notion and using the same or similar thing like an object of measurements. When only the elements of localized algebra can be an object of measurement this cannot be an obstacle for s.\,o. to have an idea or even a theoretical model of localization.
To use formula (\ref{KMS2}) like an instrument of prediction I need to know by measurement the quantum algebra on a restricted Cauchy plain in Rindler wedge. But to understand the meaning of (\ref{KMS2}) I need to know the complete net of algebras \textbf{only as an abstract notion, without any ability to measure something behind the horizon}.

Unruh--Rindler model is not an S-matrix problem. There isn't such thing like in- and out-vacuum there, only a KMS-state.  An observer recognizes this state by the spectrum of detected quanta. That's all.

According to the global state, an important duty appears in the case of curved space-time~-- the theorist (not the observer) will have to verify whether the global invariant state is a Hadamard state or not. If it isn't then QFT on the classical gravitation background is not self-consistent.
The next question is coming:

Can we check whether the global invariant state is physically meaningful without building the global Wightman function?

Yes, we can check whether or not a $\beta$-KMS Wightman function is a Hadamard type solution.



KMS-state is an extremely symmetrical notion. The formula (\ref{KMS2}) is a demonstration, what can be extracted only by symmetry before knowing the details of the dynamics.

\subsubsection{Vacua in this idealization}
\begin{description}
 \item[Unruh vacuum] is related to a normal mode expansion in stationary (or static) region. It is a pure ground state of the quantum field when the algebra is treated there as separately built one. Near the horizon they look like monochromatic waves in Eddington--Finkelstein coordinates.
  \item[Boulware vacuum] is a global isometrical invariant vacuum, built by the extended normal mode expansion of two-point distribution.
  \item[Kruskal vacuum] is related to Kruskal mode expansion of a global quantum field. Kruskal modes are not attached to the symmetry, but near the horizon, they look like monochromatic waves in Kruskal coordinates. (They correspond to the Lorentzian isotropic coordinates in Minkowski space.)
  \item[Hartle--Hawking state] When the local quantum algebra in the stationary region is treated as a restriction of the full quantum algebra the global vacuum(state) thermalizes to a KMS-state called Hartle--Hawking state.
\end{description}

\subsection {The picture describing the horizon as a source of radiation is misleading}
Here we are going to discuss the physical meaning of particles creation. Is there an emission or only creation?

Outside the algebraic framework, the quantum BH physics has its important results is worth mentioning them.
Boulware's work \cite{B1} is a key. A global quantization is accomplished (realized)
by building of the quantum Green function expanded in \emph{eigenmodes}\footnote{\ This is an extension of the notion `mode' coming from the physics of wave processes. In gravitational physics `eigenmodes' is simply the eigenfunctions of a Killing vector even when the last becomes space-like and the corresponding eigenvalues cannot be related to frequencies.}.
Quantization in Rindler coordinates has used as a guide to quantization in Schwarzschild coordinates. For 16 combination of location of $\x,\,\x'$ in the four regions Boulware patched the regional solutions by the wave packet technics.
When at least one of the points lies in a static region, the convergency of the solution at the spatial infinity is taken as a standard boundary condition (BC) for \textbf{every} Green function.
When at least one of the points lies in a region behind the horizon, so called \emph{positive frequency boundary condition}\footnote{\ It means that the Fourier expansion of $G_{\!F}$ must contain only positive (negative) frequency Lorentzian modes at the infinite future (past). In the standard courses in QFT on Minkowski space this spectral condition
 is encoded in the choice of contour going round to the poles in the complex plane instead as an explicit BC on the plain of infinite future/past.} has to be taken as a quantum spectral condition. This BC specifies the particular Green function as a Feynman propagator.
 In the internal areas, the hyperboloids at the infinite distance from Rindler edge $UV=const\to+\infty$ coincide with the space-like plains at the infinite future/past in Lorentzian coordinates. In this way the quantization in Lorentzian coordinates turns out to be a supporting structure for the determining of positive frequency BC for the case of Rindler mode expansion. A direct check shows a coincidence of extended Rindler mode expansion of $G_{\!F}$ with Lorentzian mode expansion. In other words boost invariant global state(vacuum) is equivalent to Minkowski state(vacuum).

Such a parallel structure is missing in the case of Schwarzschild space-time, where there is a real singularity at a finite distance from the bifurcation sphere, instead surfaces attached to the symmetry at the infinite future/past. So no way to copy the result of quantization in Rindler coordinates to the case of Schwarzschild geometry. That's way the task of fixing the correct positive frequency BC seems to be the most nontrivial point in \cite{B1}.

Although Boulware \cite{B1} does not consider thermalization of the field, he obtains the following important results:
\begin{itemize}\bf
  \item Any quantum emission is not possible from the interior of an eternal Schwarzschild  BH.
  \item The global vacuum in (maximally extended) Schwarzschild space-time is stable.
\end{itemize}
In \cite{B2} the research continue with fermionic fields.

According to Hawking's concept, the author makes a distinction between maximally extended Schwarzschild space-time and the case of BH formed by collapse. \textbf{The key moment of Boulware's results consist in an emphasizing the role of the BH formation for Hawking radiation.}

How to make compatible the fact of missing radiation for the present level of idealization with the prediction of thermal quanta even at the spatial infinity?

A little precondition for misunderstanding could come from the concept of ordinary black body. But a comparing between the last and an eternal black hole shows drastically different physical pictures.\\
\noindent\begin{minipage}{\textwidth}%
\begin{multicols}{2}%
\phantom.\hfill\textbf{Ordinary black body:}\hfill\phantom.
\noindent\begin{enumerate}\itemsep=0ex\small
  \item Thermodynamic medium is inside the compact boundary
  \item In the case of light, where there isn't any interaction between quanta, as it is in the case of classical ideal gas, we need a picture of interaction with the boundary~-- an incoherent absorption and reemission leading to thermalization.
  \item Black-body emission cannot be excluded by any choice of the state of motion of the observer.
  \item The temperature has got an actual character like in the case of classical ideal gas.
  \item Relativity of simultaneity is negligible, so the equilibrium leads to one and the same local temperature.
  \item There isn't any quantum entanglement.
\end{enumerate}
\columnbreak

\phantom.\hfill\textbf{Eternal black hole:}\hfill\phantom.
\noindent\begin{enumerate}\itemsep=0ex\small
  \item Thermodynamic medium is in the BH exterior, so it is an example of infinite volume system.
  \item No need of implying any kind of process of interaction with the boundary (horizon).
  \item Particle creation can be excluded with a proper choice of observer.
  \item The temperature is a potential notion. It can be actualized as a local temperature by a particle detector.
  \item Nonsynchronous clocks attached to the time symmetry manifested in different local temperatures is an essential relativistic effect.
   \item The quantum field in BH exterior is highly entangled with that in the interior.
\end{enumerate}
%
\end{multicols}
\end{minipage}\vspace{3ex}\\
\textbf{The BH interior cannot be a model of thermodynamic system} not because an observer living there cannot send to the exterior the results of his measurement\footnote{\ The common concept of local measurement excludes the information interchange.}, but because the interior cannot imply a model of an equilibrium system. If the global algebra has been localized in a black or white hole the free field dynamics there cannot lead to an attached to symmetry (isometry) construction of algebraic evolution. That's why \textbf{any KMS-state cannot exist for a region where Killing vector $\Kv$ becomes space-like}.

One more time we face a similar modernistic situation like in the case of BH mass where a compact object is characterized by the properties of its exterior. And again (in the light of Chandrasekhar's statement) we upgrade the concept of perfectness of a compact object equipping it with a thermodynamic characteristic based only on the concept of LQFT and the space-time structure without implying any process, even any interaction with the boundary (the horizon). For this reason the exterior is physically simpler than the statistic model of classical ideal gas although the complication of a large scale medium leading to different local temperatures.
Inequality of local temperature is related to the nonexistence of synchronized clocks, attached to the symmetry. The temperature which we attribute to the thermodynamic system (as its characteristics) is a local one, attache to a prefer integral curve of the Killing vector $\Kv$. Its proper time becomes modular group parameter.
In the contrast to black body, the Killing horizon is a very special boundary~-- a geometric boundary. As a geometric object, it is something absolute. As a boundary of a thermodynamic-media it is a context-dependent structure. We cannot speak about the temperature until we measure it. We always have a context-dependent situation in every attempt for actualizing the temperature.

We cannot ask whether the quanta are really there before detecting them, in the same way, we cannot look at the BH temperature as we have used to do this with the ordinary black body.

The interaction as a feature of an ordinary black body in the case of BH exterior or Rindler wedge is replaced by a more subtle thing~-- the field localization.


\setlength\itemindent{-1.3em}
\begin{itemize}\bf
  \item KMS-state is a way to equip a BH with a characteristic temperature without involving any process.
\end{itemize}

There is a widely propagated picture in which the horizon, more exactly near the horizon region, is represented as some kind of `source' of thermal emission. This makes an impression that there is some real material source of quanta located near the horizon.
It's important to be fully aware of whether we work with a local covariant notion of source like a 4D current density in space-time or following the physical intuition about happening there, we involve noncovariant objects which paint a relative picture.
In Boulware's research, the model of source $j(\x')$  is a completely covariant, state independent notion.
$$\psi(\x)=\int G_{\!F}(\x,\x')\,j(\x')\,d^4x'$$
\ \\
An intuitive picture of Hawking radiation involves the creation of virtual particle-antiparticle pairs in the vicinity of the black hole horizon. Could we apply such kind of picture in the case of Unruh effect in Rindler wedge?
We have to be careful with some imaginations because \textbf{in every attempt to use closed to our intuition ideas we take a risk of working with noncovariant objects, therefore to paint a relative picture finding it as an absolute such.}
\begin{itemize}\bf
  \item The relative character of the existence of particles is a difficulty in the interpretation of the behavior of a quantum field, but not a defect of the formal system of notions in QFT. Just this kind relativity makes QFT compatible with the principle of equivalence in GR.
\end{itemize}
\paragraph{The horizon is invisible for free falling observer.}
A free falling observer cannot determine the moment of crossing the horizon by physical measurements since the last is not a source of particles in a covariant sense.
At this moment he is just like a uniformly moving observer in Minkowski space who crosses the horizon attached to some accelerated observer.
The local quantum algebra for such observer is not restricted to BH exterior. His state of motion is not attached to the space-time symmetry except at the spatial infinity. So there isn't a way the quantum vacuum to be seen as a thermal bath.
This is the compatibility with the principle of equivalence in gravitational physics. Otherwise it will be burned by quanta with huge blue shifted energies.\vspace{2ex}

\textsl{``Our fundamental theories of matter are field, not particle theories. Particles can however be a useful concept in that the interaction of the field with localised systems can mimic the behaviour of particles. At all times one must however remember that the fundamental theory is not a particle theory, and that describing the system in terms of particles may be misleading.''}\vspace{-1ex}
\begin{flushright}%
William Unruh \cite[\!p.109]{U1}
\end{flushright}

\paragraph{Infinite high local temperature.}
It's hard to imagine how an extremely low even zero vacuum expectation values of stress-energy tensor may correspond to an infinite local temperature. But this contradiction with our physical intuition finds a natural solution: The local temperature, in fact, is not an actual but much more a potential quality of a localized quantum field. Don't think about the quanta as a something (that) is really there. (The legacy of quantum mechanic can help.) To actualize such an extremely high temperature we need a detector moving with an extremely high acceleration. Independently how small is the mass of the detector, an extremely large quantity of energy must be spent to support preserving the state of motion. A huge amount of this spending energy is the total energy of quanta born at the moments of detecting.
It is impossible to hold an observer in a position with infinite temperature, so no way to actualize the last.
Unruh shares this kind of understanding of local temperature \cite{U},\cite[\!p.7]{W3}.

Supposing a radiation is a way to attribute statute of a real existence to the field quanta.
\paragraph{Incompatibility with the integral energy conservation law.} When there is a time-like Killing vector field a local conservation of 4D density of action leads to a conservation of integral energy.
If there is an energy transfer (radiation) from the horizon to the spatial infinity, either conservation low is violated or an eternal BH turns out to be an infinite reservoir of energy despite its finite mass.
\paragraph{Incompatibility with time-reversibility of initial idealization.}
The radiation implies a time-reversibility as in classical as well in quantum physics.
In an eternal BH just like in Rindler wedge, there isn't such thing like a quant before detecting who travel from a place with infinite high temperature to the place with a finite such. The restricted field realized itself as a quant energy just in the moment of detection. One who against this should suppose an energy transfer from the horizon to the spatial remoteness even when there isn't any detector there. But in this case, it will face with a puzzle~-- where the time irreversibility came from. The algebraic evolution (\ref{AlgEv}) leading to KMS-state is time-reversible. It cannot bring something like a radiation boundary condition. Particle number operator cannot be taken as an argument for radiation BC. So its nonzero expectation value cannot be an argument for emission.
We always have an Unruh effect in the idealization of eternal BH even at the spatial infinity, but with a Hawking temperature there.
\paragraph{Incompatibility with ordinary radiation boundary problem.}
If s.o. tries to consider a particle emission in the context of a radiation boundary problem the model independence of the spectrum shape of outgoing particles creation will be lost.
$$[\partial_U\partial_V+W_l(U,V)]\,\Psi_l(U,V)=0\,,\quad W_l(0,V)=W_l(U,0)=0$$
The metric in BH exterior is encoded in the effective potentials $W_l(U,V)$, the mass of the field is also included there, and the normal mode expansion of the solution does not give a Planck spectrum. The best demonstration is realized on the metrics allowing exact solutions of KG equation.
In the case of Unruh effect in Rindler wedge the effective potentials are exponentially growing with respect to the distance from horizon, so there isn't outgoing solution at the spatial infinity. Unruh effect in Rindler wedge is the best demonstration how particles creation is possible without energy transfer from the horizon to the accelerated detector.
\begin{itemize}\bf
  \item A bifurcation Killing horizon in curve space-time is an as real source of quanta as Rindler horizon is. In both cases nothing will happen without a detector.
\end{itemize}
 \textsl{``It is difficult to imagine how this state could naturally occur as a result of any physical process.\label{Wald1} Thus, it should be emphasized that the existence of the Hartle--Hawking vacuum in extended Schwarzschild spacetime does not provide a valid argument that a physical Schwarzschild black hole (produced by gravitational collapse) would radiate thermally. Such an argument, however, is provided by the original Hawking derivation. Conversely, our proof of nonexistence of a similar state in (extended) Kerr spacetime does not affect the validity of previously derived results for radiation by Kerr black holes formed by gravitational collapse.''}\quad/ R. Wald \cite{W1}/\vspace{1ex}

An interesting result:  a state with a finite number of particles loaded over the horizon gives a zero contribution to the particle number that will detect a remote observer \cite{D0}.

The particle counting by a remote detector has not be taken as objects travelled through space.

It has to emphasize that the detector response depends on the coupling function describing the interaction between the detector field and quantum system.
In the cases of a finite time window, it is possible Unruh type detector to account particles even during uniform motion in Minkowski space in \cite{SP}.
Then, the predicted counting due to the modulation caused by the shape of the coupling function.

Nevertheless, \textbf{the field localization in the Rindler wedge is that fully coordinateles and state independent content of the Unruh effect, which is beyond the particular model of the detector}.

\subsection{The temperature as a scaling limit}


The main advantage of LQFT consists in the ability to build a notion of characteristic temperature on the horizon itself, as a scaling limit of two-point distribution based on the Haag--Narnhofer--Stain's prescription, suggested for the first time in \cite{HNS}.
 According to PLD, in the case of linear QFT, two-point distribution near the singularity can be considered as a distribution over the tangential space $T_{\x}\times T_{\x}$. For determining the exact scaling factor, the three authors introduce \textit{principle of local stability} (PLS), which in fact is a local version of the quantum spectral condition (positivity of the energy). According to PLS two-point distribution in the momentum space has to have a support in the forward light cone

\begin{equation}\label{SL}
\lim_{\lambda\to0_+}\lambda^2{\cal W}(\x+\lambda\mathbf z_1,\x+\lambda\mathbf z_2)=\frac1{4\pi^2}\frac1{g_{\mu\nu}(\x)z^\mu z^\nu}\:,
\quad\begin{array}{|l}
    \mathbf z_{(k)}:=z_{(k)}^\mu\partial_\mu|_{\x}\,,\;k=1,2\\
    \mathbf z:=\mathbf z_2-\mathbf z_1
    \end{array}
\end{equation}

Haag, Narnhofer and Stain derive the temperature for the cases of Rindler wedge and a static BH by taking the $\beta$-KMS Wightman function.

Later the scaling limit concept reaches a completion in \cite{FH1}.

Further development of the scaling limit is a formulation of \emph{Quantum  Equivalence Principle} (QEP) \cite{Hs}.
HNS prescription is written in the context of a local inertial reference frame by constructed a one-parametric diffeomorphism scaling a point $P_1$ into $P_\lambda$ ($\lambda\in[0,1]$) so that $P_0$ is a point corresponding to the local coordinate center.

There are two criticisms in \cite{Hs} addressed to \cite{HNS} against the realization of prescription. The first one related to the treatment of the two-point distribution as an ordinary function instead as a distribution. The second one against to the fixing one of the two points at the bifurcation surface ($U=V=0$) instead of a more general position on the future horizon ($U=0,\;V>0$). For the idealization of an entire BH, the fixed on the bifurcation surface point is not a restriction of generality, since the state is invariant under the action of the isometry group, so the point at the bifurcation surface always could be pushed forward to a general position on the future horizon.

The process of BH formation implies an asymptotic time symmetry so the point has to be located at a position corresponding to the asymptotic symmetry and it cannot be pull back near the moment of the formation when the symmetry is missing.
Authors of \cite{HNS} don't pretend to describe the evaporation. They work in an idealization compatible with the algebraic notion of KMS-state.

Hessling associates Hawking effect with the process of a formation where the bifurcation surface is missing in the geometric picture. His second criticism is unreasonable because he proceeds from the other idealization.
From the other side, a general location on the future horizon may correspond to an event when the word line of a free falling observer intersects the horizon, which is important for a treatment of the quantum principle of equivalence in the context of scaling limit.

Scaling limit predicts the leading singularity term of the two-point distribution of (\ref{H-state}) and it also predicts that the mass of the scalar field can be found only in the rest terms.

When we pull up to the horizon from the outside we can take $\beta$-KMS two-point distribution instead of the two-point distribution of the global invariant state.

{\small In vacuum-vacuum transformation the mass independence of the temperature originates from the mass independence of both~-- the Kruskal mode decomposition of the field, (which represents the quantum field in whole global Cauchy region) and Unruh mode decomposition (which represents the localized in right stationary region field)~-- because both kind decompositions are taken in the vicinity of the horizon, where the information about the mass disappears.}

In the case of the extremal BH, where the Killing horizon does not of the bifurcation type, there aren't there Kruskal like coordinates, the corresponding modes and associated to them Kruskal vacuum, so there isn't a ground for applying a vacuum-vacuum transformation to prove the validity of the famous formula $T=\sg/2\pi$ when $\sg=0$. Then using the scaling limit can be a decision.

Moretti consider the extremal BH \cite{M},

There are few aspects that will have to be mentioned because they are achievements of the LQFT.

In the middle of 90-th physicists studying the BH models in the algebraic framework have not restricted the exploration only to the globally hyperbolic manifold.

Not only Reissner--Nordstr\"om but every extremal compact Killing horizon in a stationary space-time is actually a global Cauchy horizon, so from general considerations it can expect a zero characteristic temperature. Why?
The full quantum algebra is located in the BH exterior so there isn't additional algebraic restriction leading to thermalization. The region of algebraic evolution (\ref{AlgEv}) occupies the total Cauchy region. Unless of a possible nonexistence of a ground state, I can't see any other shortcoming in this argument. To reject the nonexistence, we refer to \cite{Str}, which is a latter result than the commented work. The algebraic aspect of an eternal extremal BH express in a degeneration of the TT-modular structure, so outside the scaling limit prescription it is natural to expect an extremal KMS-state.

In the end of \cite{M}, the result is compared with the derived in the \cite{AHL} formula, where the zero temperature is a condition for regularizability of the vacuum expectation of the stress-energy tensor of an extremal R-N BH.
It has to be denoted that there isn't a logical connection between the two results. The algebraic notion for KMS-state bases on Reeh--Schlieder's theorem \cite{Str}, while the consideration of regularizability has got a completely different origin.
It is not clear if such a regularizability will be valid for a compact extremal Killing horizon for other gravitation solutions.

In the same way the result in \cite{Str} does not in a contradiction with the cases of nonexistence pointed out in \cite{KW}, because it indicates a general algebraic notion of ground- or KMS-state, which could be physically meaningless.
(Hadamard condition is taken into account in \cite{KW}.)

Nevertheless, since the extremal BH is a very special case it raises requirements for a better precision of HNS scaling limit prescription.
It always is beneficially for a better understanding to intersect a result from different approaches (the TT-modular structure and the HNS scaling limit).

What is important to know, \textbf{neither the scaling limit prescription nor the vacuum-vacuum transformations can verify whether the global state restricted to a KMS-state is really a Hadamard state or not, so there is a risk the formal temperature we derive to turn out be physically meaningless.}
\begin{itemize}\bf
\item By the understanding of the characteristic temperature as a scaling limit of two-point $\beta$-correlation function we have a local notion of characteristic temperature\footnote{\ It has not to be mixed up with the local temperature, determined by thermal spectrum of the detected quanta.} as a constant at any point of the stationary region as well as on the Killing horizon.
\end{itemize}

\subsection{The case of asymptotically nonflat static space-time with one bifurcation Killing horizon}
A completion of the theme of model independence of temperature requires a discussion about this case.
There ware people on the internet who puzzled about how to determine Hawking temperature in the presence of a cosmological horizon. Then the region of stationarity is bounded by two bifurcation Killing horizons with two different surface gravities, so the understanding of characteristic temperature as a result of Wick rotation becomes misleading. According to the theorem of uniqueness \cite{KW}, KMS-state cannot exist for Schwarzschild--de'Sitter BH.

How about the asymptotically nonflat static space-time with one bifurcation Killing horizon?

It is important to emphasize that LQFT doesn't fix the final value of temperature since the last requires a fixed etalon of time. One can easily see when a coordinate transformation changes the time scale, the surface gravity will also change.\vspace{-1ex}
$$t\to\lambda t\quad\Rightarrow\quad g_{00}\to\lambda^{-2}g_{00}\quad\Rightarrow\quad\sg\to\lambda^{-1}\sg\,.
$$
The choice of the etalon of time means the existing of a preferred place, where the Killing vector can be normalized. The simplest geometrical structure related to a clock is the\textit{ photonic orbit}~-- the place where the periodicity in space lead to a periodicity in time. So \textbf{the characteristic temperature always can be set as a local temperature on the photonic orbit}.\vspace{-1ex}
$$T_{loc}=\frac{T_{ph}}{\sqrt{-\mathbf g(\Kv',\Kv')}}$$
 This definition is also applicable to the asymptotically nonflat space-time where the norm of time-like Killing vector infinitely grows to the spatial infinity. In this case, however the time etalon will be chosen the characteristic temperature will be gravitationally red-shifted to the zero local temperature at the spatial infinity. So any creation of thermal quanta does not possible there.
\begin{itemize}\bf
  \item The case of the asymptotic nonflat space-time teaches us to accept the characteristic temperature of a KMS-state as a condition for the existence of a finite nonzero local temperature at the finite distance from the horizon rather than as a local temperature at the spatial infinity. This determines the localized field as a thermal bath.
\end{itemize}


In the case of asymptotically flat space-time, the time-like Killing vector is preferably normalized at the spatial infinity instead on the photonic orbit. The infinity is the best reference position for comparing of two BH.
 Another important reason for this choice is a relation between the surface curvature and the mass of the BH. The mass is a completely global notion that (in the contrast to temperature) can only be defined at the spatial infinity and only for the case of asymptotically flat space-time.

\textbf{Unless the cases when KMS-state does not exist at all, the asymptotically nonflat static space-time with one bifurcation horizon is an example for a well-defined notion of temperature when an adequate definition of mass of BH doesn't exist.}

\begin{center}*\qquad*\qquad*\end{center}

\begin{itemize}\bf
    \item A pure algebraic notion KMS-state give the best answer to the question:\\ Why the different metrics lead to the well-known quantum statistical spectra?
    \item It is an exact result leading to a correct notion of temperature without introducing of any kind process.
\end{itemize}

\subsection{Entropy and temperature}
After the obtaining of temperature as a characteristic of a proper notion of equilibrium state, the entropy can be introduced only as a thermodynamically conjugated to it quantity, e.\,c. it remains a result of a formal redistribution of fundamental constants in the first term of Smar's formula. The reason is that for infinite volume systems, the operator of algebraic evolution (\ref{AlgEv}) doesn't belong to the trace-class. The density matrix is missing to define an entanglement entropy.
So we may say that the \textbf{LQFT on a classical gravitational background cannot offer an entropy without an additional assumption allowing finite number degrees of freedom}. That's why the LQFT on a classical gravitational background cannot provide validity of Bekenstein's conjecture based on first principles.

\begin{itemize}\bf
\item The KMS-state is the purest probably unique connection between the local QFT and thermodynamics without involving additional assumptions.
\end{itemize}

Most of the scientists share the view that the deeper connection between BH thermodynamics and quantum physics has to be searched outside the quasi-classical approximation.

In the case of asymptotically nonflat space-times, there is a temperature but there isn't even a formal way for deriving the entropy since Smar's formula is missing. This emphasizes the need of searching an independent on the temperature definition.

Quantum physics in the presence of horizon because of its extreme nontriviality is a highly beneficial area for new hypotheses.

According to Suskind's point of view all degrees of freedom of a quantum field are stuck on the horizon surface but from inside forming a firewall, while according to LQFT the equilibrium system is realized outside.
(By the way, Fredenhagen notes that the algebraic net for the case of standard treatment of horizon corresponds to von Neumann algebras with type III$_1$ factor in the contrast to firewall treatment where it should be type I \cite{F0}.)

Recently four enthusiasts offer a Gedankenexperiment \cite{AMPS}, which settles a contradiction between the basic paradigms in the fundamental physics. The result
One of the following three statements must be rejected:
\begin{itemize}
  \item Unitary of evolution
  \item The present status of the QFT
  \item The principle of the equivalence in gravitation
\end{itemize}
It is constructed a situation violating \textit{the monogamy of the entanglement states}.
An important feature of the paper is adopting finite number degrees of freedom at a very beginning.

More serious ground for a derivation of entropy from first principles can be sought in the model of evaporating black holes.

\section{The collapse as an influence initiating a quantum transition}
For completeness of the parallel between Hawking effect and Unruh effect in Minkowski space with respect to idealizations we also need to spend a time for the following case.
\subsection{Appearing in Minkowski space horizon}
It can be simulated by the changing of the state of motion of an observable. It could be taken as an analogue of a collapse. A formal abstract observable is not enough, the detector is important.
So if we suppose that the detector can be switch off/on, what is the next its most important property? The measurements in working state have not dependent on the state of motion before switching on and after switching off. That's it.
Let the state of motion before the moment $t_1=0$ be a rest. After that moment the detector starts moving with a constant acceleration in a working state till the moment $t_2$ when  is switched off.
According to out assumption this picture can be replaced by another one when the piece of word line between $t_1$ and $t_2$ belongs to a piece of integral curve of boost field in the Rindler wedge.
This is another point of view when near the horizon area cannot be taken as a source because the picture of appearing horizon implies a time delay of the arriving quanta and the counting will be different.

Although the regions of the field localizations are theoretically different the results of measurements have to lead to same results with an accuracy to a statistical error. Obviously the effective localization for the combined system quantum field--working detector is determined by the detector state of motion.
So in the case of Minkowski space the situation with appearing horizon does not require a separate treatment.

Such a kind of logic cannot be applied to the case of BH, since the horizon is a characteristic of the space-time.

\ \\

In the idealization of black hole produced by a collapse we also have an algebraic localization but it is not of a modular type\footnote{\ The modular conjugation is not possible.}. The asymptotic symmetry in a far future is not enough for it. The key missing property is the time-reversibility.
A simple global notion of equilibrium requires a global symmetry in time, so there aren't mirror-images ${\cal L}$ and ${cal P}$. This is the main disadvantage of KMS-state.
That's why, for this level of idealization \textbf{KMS-state is not an applicable notion for describing a BH evaporation} although the evaporation of massive BH is an adiabatical (quasi equilibrium) process.
So vacuum-vacuum transformation remains as an alternative.

To acquire an intuition what happen, when the collapse closes a horizon we shall use the model of 1D parametric oscillator as a pedagogical skill offered in \cite{J}.

\subsection{Parametric excitations}
Let's start with a classical dynamic system with a Hamiltonian and equation of motion:
\begin{equation}\label{POsc1}
H(p,q,t)=\frac{p^2}{2m}+\frac{k(t)}2x^2\;\;\longrightarrow\;\;\ddot x(t)+\omega(t)^2\,x(t)=0\;,\quad\omega(t):=\sqrt{k(t)/m}>0\,,
\end{equation}
that will be quantized by the standard prescription:
\begin{equation}\label{POsc0}
x(t)=\overline{x(t)}\;\;\stackrel{Q}\longrightarrow\;\;\left\{\begin{array}{c}
                          \hat x(t)={\hat x(t)}^\dag\\
                         {[\hat x(t),\dot{\hat x}(t)]}=i\hbar/m
                        \end{array}\right.
\end{equation}
Let's formally set $$\hat x(t):=f(t)\,\hat a+\overline{f(t)}\,\hat a^\dag\,,$$
where the constant operator $\hat a$ is chosen dimensionless. So after replacing this in (\ref{POsc0}), the last takes a form
$$(f(t)\dot{\bar f}(t)-\dot f(t)\bar f(t))[\hat a,\hat a^\dag]=i\hbar/m$$

The expression in front of the commutator is Wronskian $W[f,\bar f](t)$.

The following dimensionless structure can be defined
\begin{equation}\label{POsc3}
\footnote{\ This symbol was chosen to distinguish this inner product from a scalar product in Hilbert space.

}\,\FSP{f,g}:=\frac{im}\hbar(\bar f\dot g-\dot{\bar f}g)
\end{equation}
If $f(t)$ and $g(t)$ are partial solutions of the equation, then the complex conjugated functions $\bar f(t)$ и $\bar g(t)$  are also solutions. Since first derivative is missing in (\ref{POsc1}), the Wronskian is an integral of motion, which make possible $\FSP{f,g}$ to be considered as well defined inner product in the complex linear space of solutions. Besides the linearity the following properties we have:
$$\FSP{f,g}=\overline{\FSP{g,f}}\;,\quad\FSP{f,f}=-\FSP{\bar f,\bar f}\;,\quad\FSP{f,\bar f}=0$$
The case $g=f$ give $\FSP{f,f}\in\mathbb R$, a  not positive defined norm of vector.

In that way the commutation relation (\ref{POsc0}) is reduced to
\begin{equation}\label{POsc4}
  \left[\hat a,\hat a^\dag\right]=\hat1\;\Leftrightarrow\;\FSP{f,f}=1 \textit{ (normalized solution)}
\end{equation}

For Hilbert space  $\cal H$ in which acts the operator $\hat x(t)$, the state $|0\rangle$ takes its standard definition:
$$\hat a|0\rangle=0\quad\wedge\quad\langle0|0\rangle=1\,,$$
a condition allowing an association of $\cal H$ with a space of excitations.

The evolution of the state $|0\rangle$ in Heisenberg representation takes the form\\
$$\psi(t)=\exp(-i\hat H(t)t/\hbar)|0\rangle$$
$$\hat H(t)=\frac12m\left[(\dot f^2+\omega(t)^2f^2)\hat a^2+(\,\dot{\bar{\!f}}{}^2+\omega(t)^2\bar{\!f}{}^2){\hat{a}^{\!\dag}}{\,}^2+(|\dot f|^2+\omega(t)^2|f|^2)(2\hat a^\dag\hat a+\hat1)\right]$$
\begin{equation}\label{POsc5}
  \hat H(t)|0\rangle=\frac12m\left[(\,\dot{\bar{\!f}}{}^2+\omega(t)^2\,\bar{\!f}{}^2){\hat{a}^{\!\dag}}{\,}^2+(|\dot f|^2+\omega(t)^2|f|^2)\right]|0\rangle
\end{equation}
Since $\langle0|{\hat{a}^{\!\dag}}{\,}^2|0\rangle=0$, the vector ${\hat{a}^{\!\dag}}{\,}^2|0\rangle$ could not be collinear to $|0\rangle$, so the first term in (\ref{POsc5}) must vanish $|0\rangle$ to become an eigenvector of $\hat H(t)$. The condition $\dot f(t)=\pm i\omega(t)f(t)$ could be made compatible with the equation $\ddot f(t)+\omega(t)^2\,f(t)=0$ in the case of existence of an asymptotic behaviour with constant frequencies.

$$\ddot f(t)+\omega(t)^2\,f(t)=0\quad\Rightarrow\quad
f(t)=\left\{\begin{array}{ll}
f_-(t)=A_-e^{-i\omega_{in}t}+B_-e^{i\omega_{in}t}\,,&\textrm{when }t\to-\infty\\
f_+(t)=A_+e^{-i\omega_{out}t}+B_+e^{i\omega_{out}t}\,,&\textrm{when }t\to+\infty
\end{array}\right.$$
In the other words the dynamical system has to possess asymptotic time symmetries as it is in the case of BH formation.

\emph{The normalized positive frequency asymptotic solutions} in far past~-- $f_{in}$ and far future~-- $f_{out}$, are \emph{(uniquely)} determined as
eigenfunctions of the operator of asymptotic symmetry $d/dt$ with a norm equal to 1.
$$\left.\begin{array}{|l}
    \dot f_{k}:=\pm i\omega_{k}f_{k}\\
    \FSP{f_{k},f_{k}}:=1
  \end{array}\right\}\quad\Rightarrow\quad
  f_{k}(t)=\sqrt{\frac\hbar{2m\omega_{k}}}\,e^{-i\omega_{k}t}\;,\;\;k:= in, out
$$

One can see looking at (\ref{POsc5}), that to be initial state $|in\rangle\!=\!|0\rangle$ an eigenvector of $\hat H(t=-\infty)$, it has to set
$$\underline{f_-(t):=f_{in}(t)}\quad\Rightarrow\quad\lim_{t\to-\infty}\hat x(t)=\sqrt{\frac\hbar{2m\omega_{in}}}(e^{-i\omega_{in}t}\hat a+e^{i\omega_{in}t}\hat a^\dag)$$
$$\lim_{t\to-\infty}\hat H(t)|in\rangle=\frac m2(|\dot f_{in}|^2+\omega_{in}^2|f_{in}|^2)|in\rangle=\frac{\hbar\omega_{in}}2|in\rangle$$

\begin{equation}
f_+(t)=\alpha f_{out}(t)+\beta\bar f_{out}(t)\neq f_{out}(t)
\end{equation}
Since  $\FSP{f,f}=\FSP{f_{in},f_{in}}=\FSP{\alpha f_{out}+\beta\bar f_{out},\alpha f_{out}+\beta\bar f_{out}}=1$, \ then
\begin{equation}
|\alpha|^2-|\beta|^2=1\,,
\end{equation}
where \ $\alpha$ and $\beta$ are \emph{Bogolyubov's coefficients}. They depend on $\omega(t)$.

A formal set
$$\hat a:=\FSP{f_{in},\hat x}\;\;,\quad\hat b:=\FSP{f_{out},\hat x}\;,
$$
lead to the relations:
\begin{equation}
  \hat b=\alpha\,\hat a+\beta\,\hat a^\dag\quad\Rightarrow\quad [\hat b,\hat b^\dag]=\hat1\,.
\end{equation}
After defining a vacuum with respect to the far future (out-going vacuum)
$$\hat b|out\rangle=0\quad,\qquad \langle out|out\rangle=1\;,$$
we can see how the in-going vacuum looks like a state with certain number of quanta with respect to out-going vacuum
$$\langle in|\hat b^\dag\hat b|in\rangle=\langle in|(\bar\beta\,\hat a+\bar\alpha\,\hat a^\dag)(\alpha\,\hat a+\beta\,\hat a^\dag)|in\rangle=|\beta|^2\,.$$

Although $|in\rangle$, $|out\rangle$, $\hat a$, ${\hat a}^\dag$, $\hat b$, ${\hat b}^\dag$ are correct objects during the whole process, an interpretation of  $|in\rangle$ ($|out\rangle$) as a ground state can be applied only to asymptotic past (future).  An interpretation of ${\hat a}^\dag\hat a$ и ${\hat b}^\dag\hat b$ as operators of number of quanta is only possible when both commutate with $\hat H(t)$, so there isn't any good notion of quanta (particles) at an intermediate time.

To escape deal with WKB-approximation, we suppose the model of a system under a quick stress (shock).
\begin{equation}
\omega(t)^2=\omega_{in}^2+\Theta(t)(\omega_{out}^2-\omega_{in}^2)\;\;\Rightarrow\;\;f(t)=
\left\{\begin{array}{cc}
 f_{in}(t)\,&t\leqslant0\\
 \alpha f_{out}(t)+\beta \bar f_{out}(t)\,&t\geqslant0
 \end{array}\right.
\end{equation}
The Bogolyubov's coefficients can be determined by conditions patching the solutions:
\begin{subequations}
\begin{equation}
  \begin{array}{|l}
   \displaystyle\lim_{\varepsilon\to0}f(-\varepsilon)= \lim_{\varepsilon\to0}f(\varepsilon)\\
   \displaystyle\lim_{\varepsilon\to0}\dot f(-\varepsilon)= \lim_{\varepsilon\to0}\dot f(\varepsilon)
  \end{array}\;\;\Rightarrow\quad \alpha=\frac12\br{\sqrt{\frac{\omega_{out}}{\omega_{in}}}+\sqrt{\frac{\omega_{in}}{\omega_{out}}}\;}\,,\;\;\beta=\frac12\br{\sqrt{\frac{\omega_{out}}{\omega_{in}}}-\sqrt{\frac{\omega_{in}}{\omega_{out}}}\;}\,,
\end{equation}
\begin{equation}
  |\beta|^2=\frac14\br{\frac{\omega_{out}}{\omega_{in}}+\frac{\omega_{in}}{\omega_{out}}-2}>0\;\;\Leftrightarrow\;\;\omega_{in}\neq\omega_{out}\,.
\end{equation}
\end{subequations}
The considered task is the simplest model describing a creation of particles. Our aim was to emphasize the role of the quantum transition. The typical time for the collapse of a star with a solar mass is about few milliseconds, when the time of evaporation of the BH is many orders longer than the age of Universe, so the shock influence is an almost perfect approximation.

 Although vacuum-vacuum transformations in space-time look similar in both idealization, in the case of formed BH they are related to a quantum transition(excitation) in a contrast in the case of eternal BH where they aren't. The real field localization is a process
The awareness for this meaning could be lost if we mix two idealizations.

If we go back to the Wald's statement cited on p.\,\pageref{Wald1}, everyone can see a clear distinction between the two idealizations.
It is emphasized that the result for each of them cannot be an argument to other. The approaches are different.
But in a later review \cite{W4} Wald said:
\textsl{``The differences between the Unruh and Hawking effects can be seen dramatically in the case of a Kerr black hole. For the Kerr black hole, it can be shown [42] that there does not exist any globally nonsingular state of the field which is invariant under the isometries associated with the Killing horizon, i.e., there does not exist a ``Hartle--Hawking vacuum state'' on Kerr spacetime. However, there is no difficulty with the derivation of the Hawking effect for Kerr black holes, i.e., the ``Unruh vacuum state'' does exist.''}

(Unruh state in this idealization is the same as in the previous.)
When s.o. has to talk about the difference between Unruh effect and Hawking effect, it will be better to use one and the same idealization. In a contrast of the statement in \cite{W1}, here it is not clear the phrase ``Kerr black hole'' in the first sentence of the quotation \cite{W4} which idealization refers to. In the second sentence from the current context as well as from the end~-- ``Kerr spacetime''~-- we understand that it stay word about the case of an exact symmetry. But ``Kerr black holes'' in the third sentence are already formed by collapse. Such a mixture makes the statement logically untenable.
The question here is the following: If the equilibrium state doesn't exist for Kerr BH, why we cannot expect the same for a formed rotating BH, which seems as? The answer is here \cite{W2}.

The real complexity of Hawking effect is in all things that cannot be met in the model of the parametric oscillator. These are the specificities of the local QFT in curved space-time as well as the consequences leading beyond the QFT on a classical gravitational background.

What happens when the horizon appears? The Cauchy surfaces associated to the observer are shrunk after the certain moment.
The importance of the Cauchy surfaces is an essential lesson of QFT on the curve space-time as well as on the Rindler wedge.

The violation of the unitary of the evolution is closely related to the information paradox.

Wald remarks in a public talk that the violation of the unitary a happens when there is a transition from a complete system of observables to uncomplete such, which can be fully compatible to QFT.

LQFT offer an original derivation of Hawking's radiation \cite{FH2}

\ \\




\begin{itemize}\bf
        \item The main difference between the Hawking and Unruh effects is that Unruh effect always requires a detector of particles in the contrast to this, the outcome of the Hawking effect can be measured either as a radiation by a particle detector or
pure classically by variation of the gravitational field due to the evaporation.
\end{itemize}

\section{Conclusion}
Local QFT is based on rigorous notions and concepts where the interconnections between them are express by theorems. It offers a global view, an access to subtle problems, which could never expect in a style of work based only on the nonequivalent vacua.

The most important results related to Unruh and Hawking effects are traced and systemized. The significant achievements of the algebraic QFT in this context are:
\begin{enumerate}\itemsep=-0.3ex
  \item The notion of KMS-state as a most important structure characterized by a temperature when a space-time with a horizon possess a global time symmetry.
  \item Scaling limit prescription determining the temperature, as a consequence of Haag's principle of locality in the case of a linear field theory.
  \item Hadamard criterion for the state with a physical meaning, which can reject improper space-time models.
\end{enumerate}

The algebraic framework is also a key for a better distinction between the idealization of eternal black hole and the formed black hole because it shifts the focus of the attention 
from the nonequivalent vacua to the field localization.
\begin{itemize}\bf
    \item In the case of curve space-time with a bifurcation Killing horizon (eternal non-extremal black hole) the modular localization is a more fundamental theoretical phenomenon which implies the existence of local temperature rather than the existence of Hawking temperature. The last also requires asymptotic flatness for fixing a prefer notion of surface gravity.
\end{itemize}
In simple words, this means that a well defined nonzero characteristic temperature is also possible even when the local temperature at the spatial infinity becomes zero.
This may happen for some models of asymptotic nonflat space-times.

The examples of a mixture of the two idealizations \cite{HH,S,M,W4} ware presented.

Only LQFT can offer examples \cite{KW} demonstrating a missing of a smooth logical transition between the two idealizations.

Although the black hole temperature is an absolutely model independent notion, the radiation depends on space-time metric but much more depends on the particular field sector.
The dynamical aspect of the Hawking effect (radiation) face a lack of support by the algebraic QFT, where the constructive QFT offer calculated luminosities for a variety of cases \cite[\!p.409]{Wi}.

Algebraic (local) QFT offers a firm ground for research on the self-consistency of the theory in an area when the experiment is almost impossible.
There are no more admissions in it than the suggestion that, the laws verified in Earth conditions are also valid in cosmos where the influence of the space-time becomes significant.
The only additional structure, supposed as an upgrade to basic axioms and a quantum principle of the locality are Wald's axioms \cite{W2}, which simply set a self-consistent connection between QFT and classical gravity.

Methodological aspect concern in the quantum BH physics:

Physics at the present of the horizon is a remarkable demonstration of a situation, separating the concept of dynamics from the concept of evolution.
The dynamics live in a global Cauchy region~-- the region of quantization and field symplectic structure. But an attached to the symmetry notion of evolution is restricted to the stationary region of space-time.
This separation will become much deeper if someday physics finds a universally accepted way to extend the field dynamics beyond the Cauchy horizon.


Another separation we met in quantity temperature and entropy

They always are taken as a pair with respect to the (property of) existence. But the pair is broken in quantum black hole physics.
 The additional assumptions to the theory or changing the quantum paradigm is the price that the fundamental physics must pay for a recovering of this pair.
 Probably the solution to the information paradox is somewhere there.

\section{Acknowledgements}

\ \\

\ \\

\newcommand\Journ[2]{ \textsl{#1}  \textbf{#2}, }
\newcommand{\JMathPhys}[1]{\Journ{J.\,Math.\,Phys.}{#1}}
\newcommand{\PhysRep}[1]{\Journ{Phys.\,Rep.}{#1}}
\newcommand{\PhysRev}[1]{\Journ{Phys.\,Review}{#1}}
\newcommand{\PhysRevD}[1]{\Journ{Phys.\,Rev.\,D}{#1}}
\newcommand{\PhysRevLett}[1]{\Journ{Phys.\,Rev.\,Lett.}{#1}}
\newcommand{\CMPh}[1]{\Journ{Commun.\,math.\,Phys.}{#1}}
\newcommand{\CQG}[1]{\Journ{Class.\,Quant.\,Grav.}{#1}}
\newcommand\AuthTit[2]{\textbf{#1}: {\footnotesize\textsl{#2},} }
\let\RefName\refname

\renewcommand\refname{\RefName\ to classical physics and differential geometry}

\renewcommand\refname{\RefName\ to quantum field theory}


\end{document}